\documentclass[10pt,letterpaper,superscriptaddress]{article}
\usepackage{opex3}
\usepackage{widetext}
\usepackage{cite} 
\usepackage{amssymb}
\usepackage{mathrsfs}
\usepackage{amsmath}
\usepackage[normalem]{ulem}
\usepackage{color}

\begin{document}

\title{Method for predicting whispering gallery mode spectra of spherical microresonators}

\author{Jonathan M. M. Hall,$^{1,*}$ Shahraam Afshar V.,$^{1,2}$ Matthew R. Henderson,$^{1}$ Alexandre Fran\c{c}ois,$^{1}$ Tess Reynolds,$^{1}$ Nicolas Riesen$^{1}$ and Tanya M. Monro$^{1,2}$}

\address{$^1$ Institute for Photonics \& Advanced Sensing and ARC Centre for Nanoscale BioPhotonics, School of Physical Sciences, The University of Adelaide, Adelaide, SA 5005, Australia}
\vspace{-3mm}
\address{$^2$ The University of South Australia, Adelaide, SA 5000, Australia}

\email{$^*$jonathan.hall@adelaide.edu.au} 
\vspace{-3.5mm}
\homepage{http://www.adelaide.edu.au/ipas/}
\vspace{-3.5mm}
\homepage{http://cnbp.org.au}

\begin{abstract}\!\!\!\! A full three-dimensional Finite-Difference Time-Domain 
(FDTD)-based  toolkit is developed to simulate the whispering gallery modes of 
a microsphere in the vicinity of a dipole source. 
This provides a guide for experiments 
that rely 
on efficient coupling to the modes of microspheres. 
The resultant spectra are compared to those of   
analytic models used in the field. 
In contrast to the analytic models, the FDTD method 
 is able to collect flux from a variety of possible 
collection regions, such as a disk-shaped region. 
The customizability of the 
technique 
allows one to consider a variety of mode excitation scenarios, 
which are particularly useful for investigating 
novel properties of optical resonators, 
and are valuable in assessing 
the viability of a resonator for biosensing. 
\end{abstract}


\bibliographystyle{osajnl}
\bibliography{wgmref}

\section{Introduction}

Whispering gallery modes (WGMs) are produced in microresonators 
by electromagnetic waves that travel along the 
material interface. Though the WGMs are bound waves, they   
produce an evanescent field that extends beyond the 
surface of the resonator. 
WGMs represent an important optical phenomenon for sensing due to the 
sensitivity of their evanescent field to nearby entities, such as 
biomolecules, which 
break the symmetry of the field \cite{doi:10.1021/nl401633y}. 
Optical microresonators have been the subject of much recent investigation 
since their 
discovery as an important tool for biological sensing \cite{Vollmer:2002a}. 
It has been shown that microresonator WGMs can be sensitive to the presence  
of virions, animal cells, bacteria \cite{Vollmer2008591} and 
macromolecules such as proteins 
\cite{Vollmer:2002a,Boyd:01,Arnold:03,Ksendzov:05} and DNA 
\cite{pmid18036809,Vollmer:2012a}, which may be used in the development of 
label-free detection technologies \cite{Armani:2007a,Baaske2014}. 
WGMs have also been used 
for the development of high-efficiency 
optical frequency combs \cite{Liang:11}, investigating 
nonlinear optics \cite{Schliesser2010207} and quantum electrodynamics 
\cite{PhysRevA.83.063847}. 
The focus of this study is on the development of a computational 
modeling tool for investigating the properties of WGMs in microspheres. 
In particular, the advantages and limitations of the method will be discussed.

The extremely high $Q$-factors ($> 10^8$) that are possible by exciting the 
WGMs 
of a resonator often utilize a prism, 
or a tapered fiber coupled to a resonator 
\cite{Knight:97,Hossein-Zadeh:06,1674-1056-17-3-047,s100301765,6525394}. 
An alternative approach is to excite the WGMs indirectly, by using fluorescent 
dyes \cite{Francois:13}, 
 surface plasmon effects 
\cite{Armani:2007a,Min:2009a} or nanoparticle coatings such as 
quantum dots \cite{Shopova:2003a}. 
Here, the focus is on 
coupling strategies that involve 
the excitation of WGMs of microspheres 
using dipole sources. 
The Finite-Difference Time-Domain (FDTD) 
computational tool developed in this work will address 
this coupling scenario in particular. 
It should be noted, however, that 
the tool is easily extended to include 
fiber-coupled methods. 

 The use of dipole sources in the vicinity of a microresonator 
allows one to investigate the dependence 
of the optical modes on the size, shape and refractive index contrast. 
Efforts to characterize the optical properties of WGMs in resonators 
 using modeling techniques include both excitation from a 
plane-wave 
beam \cite{Fujii:2005a,Fujii:2005b,Quan:2005a}, and excitation from an electric 
dipole source \cite{PhysRevA.13.396,chance1978molecular,Gersten:80,Gersten:81,Chew:1987a,PhysRevA.38.3410,Ruppin:82,Schmidt:12,pmid24921827}. 
The dipole sources can also serve as an effective analogue for 
fluorescent 
dyes or 
embedded nanoparticles that excite 
WGMs in microspheres \cite{Francois:13}.

The Finite-Difference Time-Domain technique 
simulates the evolution of 
electromagnetic fields by discretizing a volume into a three-dimensional 
spatial 
lattice \cite{taflove1995computational}.  
Maxwell's Equations are then solved for every point on the lattice, 
for a finite number of time increments. The geometry of the resonator 
is defined for a dielectric medium, and placed in the discretized volume, with 
the edges padded with
 a field-absorbing perfectly-matched layer. 
FDTD is chosen 
specifically for its ability to incorporate effects such as material inhomogeneities, and a variety of resonator shapes, such as shells, ellipsoids or 
shape deformations to a microsphere.

The use of two-dimensional FDTD simulations to describe the resonance 
peak positions   
has been investigated in the case of microdisks 
\cite{Kuo2009}.  
However, 
the accurate prediction of the $Q$-factor of the 
resonances represents a principal challenge \cite{Little:99}, 
due to the significant dependence of the $Q$-factor 
on minute characteristics 
of the resonator, such as surface roughness, material inhomogeneities and 
microscopic deformations in the shape of the resonator
\cite{Min:2009a}. 
As a result, theoretical $Q$-factors evaluated through analytic 
models 
can be difficult to obtain experimentally  
\cite{Min:2009a,Vahala:2004}. 
The flexibility of the FDTD method is the ability to address this issue by 
incorporating geometric, material and refractive index features 
in a way that is intuitive and easy to implement. 

The FDTD method 
will be used to consider a variety of flux collection scenarios, 
source distribution, and resonator properties. 
In Section~\ref{sec:results}, an example scenario of 
a polystyrene  
microsphere in a surrounding medium of air or water is considered. 
The $Q$-factors obtained from the simulation are then compared to those 
of the Chew analytic model \cite{PhysRevA.38.3410}. 
The angular distribution of the flux density is also investigated.

Analytic models describing WGMs in the literature generally fall into 
two categories: models that provide only the positions of the TE and TM 
modes and are unable to predict the profile of the 
output spectrum \cite{Johnson:93,Teraoka:06a}, 
and models that consider the full 
behavior of the electric and magnetic fields both inside and outside 
the resonator, for an incident plane-wave (Mie scattering) 
\cite{BohrenHuffman} or 
dipole source \cite{PhysRevA.13.396,chance1978molecular,Gersten:80,Gersten:81,Chew:1987a,PhysRevA.38.3410,Ruppin:82,Schmidt:12,pmid24921827}. 
Although the simpler analytic models used in the literature are comparatively 
efficient to calculate, the profile of the 
transverse electric and magnetic (TE and TM) modes are obtained 
independently, for example, 
through geometrical arguments \cite{Johnson:93,Teraoka:06a}. 
This means neighboring modes do not interact with each other in the 
model. 
Both types of analytic model rely on the assumptions 
that the sphere is perfectly round with no surface roughness, that it
consists of a medium that is homogeneous in refractive index, 
 and that the total radiated power is collected in the limit of 
long collection times. 
The formulae for the 
power output are cumbersome, and 
must be re-derived for scenarios that consider different resonator shapes or  
inhomogeneities.

On the other hand, FDTD is easy to customize, as changes to the 
resonator geometry (spheroids, shells, etc.), 
refractive index scenario, and introducing 
material inhomogeneities can be accommodated without re-deriving the 
boundary conditions of the resonator. 
This method is able to achieve 
robust predictions for 
 preselecting specific optical properties of a given resonator 
prior to fabrication. For example, specific modes or wavelength regions 
can be reliably identified, and tailored for specific tasks 
by altering the initial configuration of the resonator. 
Furthermore, the ability to scan over a wide range of parameters 
may lead to valuable design solutions for biosensing 
that would not otherwise have been found.  

\section{FDTD method}

In this article, the three-dimensional FDTD method is simulated using the free software package, MEEP \cite{OskooiRo10}. 
Since the simulation includes the complete set of radiation and bound modes occurring in the system, the only simulation artifacts in an FDTD calculation involve discretization and finite-volume effects, and any assumptions of ideal material properties one might make. 
By making the grid size finer, 
the simulation approaches a converged result, 
 and this can be quantified in the form of a systematic uncertainty in 
the positions of the WGMs (see Section~\ref{sec:results}).

The FDTD method can lead to the possibility of accessing transient or emergent 
optical effects, as a Gaussian source is used, and 
each time step is evaluated separately. 
However, FDTD is computationally expensive. For example, time steps 
totaling a few hundred wave periods frequently require up to 100 hours of 
simulation time, for $\sim 30$ CPUs on modern supercomputers. 
Table~\ref{tab:comp} summarizes typical computing resources for a range of resolutions, using the Tizard machine at 
eResearchSA (https://www.ersa.edu.au/tizard). 
Nevertheless, for each simulation run, a large amount of data can be extracted 
by accessing the electric, magnetic and Poynting 
vector fields at each grid point. 
This allows one to map out the angular distribution of the energy density for a 
typical flux plane (see Section~\ref{sec:results}).

FDTD is also able to consider any desired position or alignment 
of dipole sources, including inhomogeneous distributions of dipoles 
placed on the surface or throughout the medium. 
The freedom to define a specific flux collection region at any point in space, 
or for any length of time,  
is an important feature for ongoing investigations into 
a range of coupling scenarios. Measuring the flux from a particular direction 
and aperture automatically 
biases the shape of the collected power 
spectrum. FDTD is therefore especially suitable for simulating 
the effect of collecting radiation through a restricted region.  

The principal drawbacks in the FDTD method are the computational intensity, 
and systematic effects due to the discretization of space. 
%NEW
The accuracy of each FDTD calculation is limited by the Yee 
cell size, which determines the grid resolution of the simulated volume. 
In the following simulations, the cell width is $20$-$30$ nm. 
A decrease in the cell size increases the computing time requirement 
to the fourth power (three spatial dimensions, and time) \cite{OskooiRo10}. 
As a result, the available hardware resources typically limit the 
improvement in accuracy of the FDTD simulations. 
This effect 
can, however, 
be incorporated into the systematic uncertainty of the calculation.

\begin{table*}[tp]
  \caption{\footnotesize{Computing resources required for a three-dimensional FDTD simulation of a $6$ $\mu$m diameter sphere excited by a dipole source with a central wavelength of $0.6$ $\mu$m. The Tizard machine at 
eResearchSA (https://www.ersa.edu.au/tizard) is used in these simulations, 
which uses AMD 6238, 2.6 GHz CPUs. 
The number of CPUs, 
 the memory (RAM), virtual memory (VM) 
and wall-time are listed for a variety of FDTD grid resolution 
values, 
$\Delta x$. 
The resolution in the frequency domain is quoted 
after being converted to wavelength, $\Delta \lambda$. 
The wall-time increases linearly with the 
flux collection time, which is held fixed at $0.6$ ps in this table. 
}}
\vspace{-6pt}
  \newcommand\T{\rule{0pt}{2.8ex}}
  \newcommand\B{\rule[-1.4ex]{0pt}{0pt}}
  \begin{center}
    \begin{tabular}{llllll}
      \hline
      \hline
      \T\B            
       $\Delta x$ (nm) & $\Delta \lambda$ (nm) & CPUs & RAM (GB) & VM (GB) & Wall-time (hrs : mins) \\
      \hline     
      \quad$33$ &\,\, $0.62$ &\,\, $24$ &\,\, $28.45$ &\,\, $34.82$ & \qquad\,\, $15:08$ \\
      \quad$30$ &\,\, $0.62$ &\,\, $24$ &\,\, $36.73$ &\,\, $43.00$ & \qquad\,\, $26:12$ \\
      \quad$29$ &\,\, $0.62$ &\,\, $24$ &\,\, $41.44$ &\,\, $47.75$ & \qquad\,\, $27:19$  \\
      \quad$27$ &\,\, $0.62$ &\,\, $24$ &\,\, $46.57$ &\,\, $52.77$ & \qquad\,\, $30:09$ \\ 
      \quad$26$ &\,\, $0.62$ &\,\, $24$ &\,\, $53.33$ &\,\, $59.61$ & \qquad\,\, $37:31$ \\
      \quad$25$ &\,\, $0.62$ &\,\, $24$ &\,\, $59.31$ &\,\, $65.62$ & \qquad\,\, $43:12$ \\
      \quad$22$ &\,\, $0.31$ &\,\, $32$ &\,\, $100.19$ &\,\, $108.02$ & \qquad\,\, $90:37$ \\ 
      \hline
    \end{tabular}    
  \end{center}
  \label{tab:comp}
\end{table*}

\begin{figure}
\centering
\includegraphics[height=160pt]{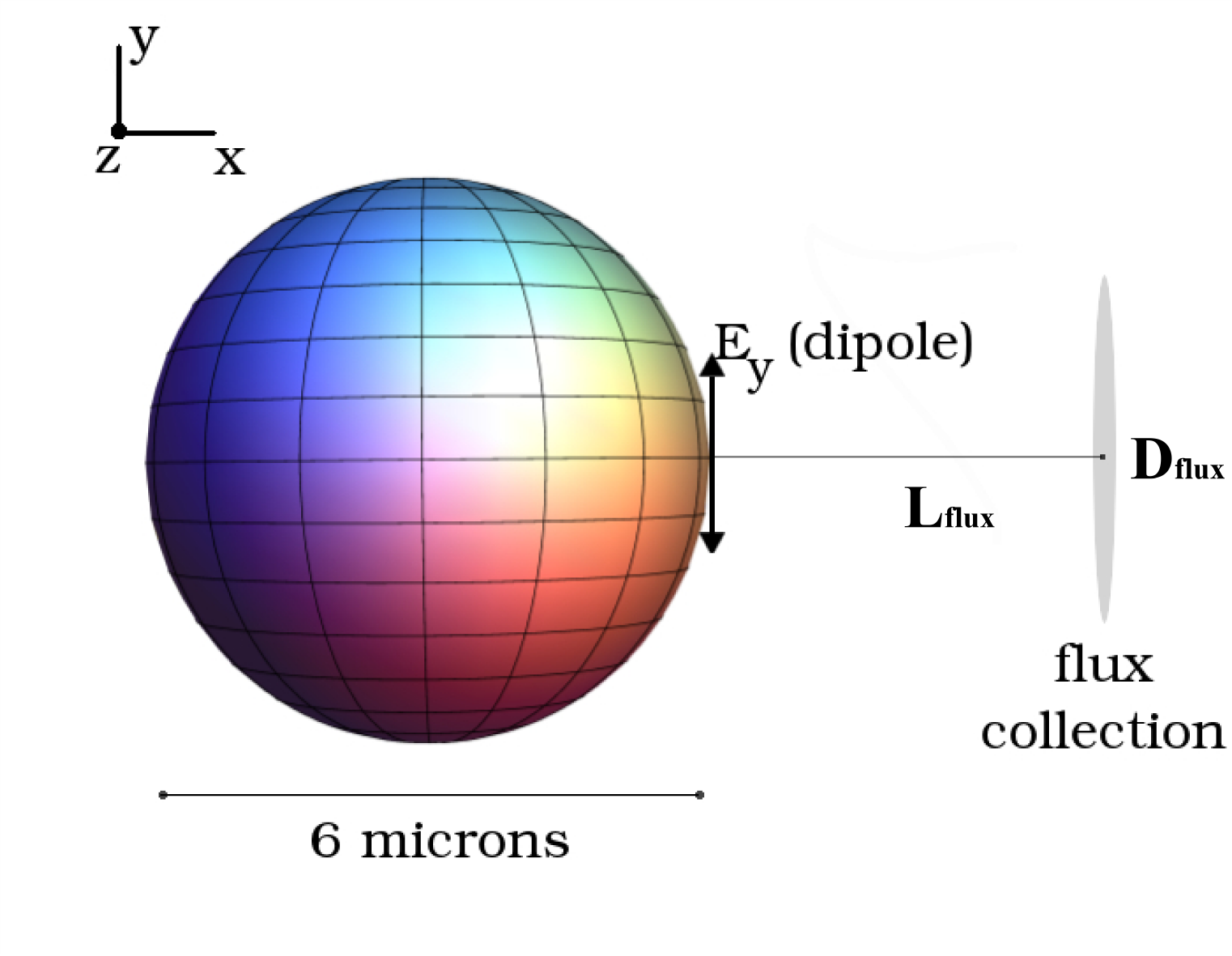}
\vspace{-3mm}
\caption{\footnotesize{ A circular flux collection region, which is offset from a $6$ $\mu$m diameter microsphere in the $x$-direction, is placed so that its normal is aligned radially outward along the same axis. In this illustration, the optical modes are excited from a tangentially-oriented electric dipole source. 
}}
\label{fig:Sphere}
\end{figure}

To generate an electromagnetic current to excite the WGMs, one or several 
electric dipole sources are placed in the vicinity of the sphere. 
Consider the example of a microsphere with a diameter of $6$ microns, as 
illustrated in Fig.~\ref{fig:Sphere}. An 
electric dipole is placed on the surface of the sphere and oriented tangentially to 
the surface,  while a circular flux collection region is defined a distance 
$L_{\mathrm{flux}}$, normal to the sphere, with diameter $D_{\mathrm{flux}}$. 
This region aggregates the flux 
in the frequency domain. 
One may vary the orientation and position of both source and collection region 
to build up a map of the coupling efficiencies to different WGMs.

The power output spectrum represents an important quantity for assessing the 
$Q$-factors and the wavelength positions of the excited WGMs. 
The total output power ($P$) in terms of wavelength 
($\lambda$) is obtained by integrating the projection of the Poynting vector 
($\mathbf{S} \equiv \mathbf{E} \times \mathbf{H}^*$) onto a  
flux region of area $A$: 
\begin{equation}
P(\lambda) = \int\!\mathbf{S}\cdot \hat{\mathbf{n}}\,\, \mathrm{d}A,  
\end{equation}
for a unit vector $\hat{\mathbf{n}}$ normal to the collection region. 
The profile of the power spectrum, plotted as a function of $\lambda$, 
will show sharp 
peaks that correspond to the positions of TE and TM WGMs. 
Note that the convention used for the TE and TM mode definitions is identical 
to that found in \cite{Johnson:93,Teraoka:06a} in which the TE (TM) mode indicates that the electric (magnetic) field is tangent to the surface of the sphere. 
The relative heights 
of the peaks indicate the coupling efficiencies to different modes, 
which is highly dependent on the geometry of the dielectric medium, the refractive 
index contrast, and also the method of mode excitation, such as dipole 
sources of different alignments, coupling to optical fibers or fluorescent dye 
coatings. 
The power spectrum  is 
normalized (in Figs.~\ref{fig:res} through \ref{fig:WGM}, and Fig.~\ref{fig:WGMang})  to the power output $P_0$
of the sources that would occur in the absence of the dielectric sphere. 
Note that the quantity $P/P_0$ is proportional to the 
emission rate ratio, $\Gamma/\Gamma_0$~\cite{PhysRevA.64.033812}. 

The corresponding cavity $Q$-factors   
can be obtained by treating the simulation output as that of an experiment, 
and measuring the central wavelength position $\lambda_0$, and the 
full-width at half-maximum, $\delta\lambda$ \cite{Vahala:2004}
\begin{equation}
Q = \frac{\lambda_0}{\delta\lambda}. 
\end{equation} 
This formula will be used in estimating the simulation $Q$-factors in 
Table~\ref{tab:qf}.

\section{Analytic models}
\label{sec:analytic}

The simplest analytic WGM model that provides the positions of the TE and TM modes 
is particularly useful for identifying the mode numbers of the WGM peaks 
that occur in the FDTD spectrum. 
While this model identifies the resonance positions, which exactly match the results of Mie scattering theory, 
 \cite{BohrenHuffman}, this approach does not give the total electric and magnetic fields including the dipole source, 
and is unable to generate the profile of the spectrum. 

The model is defined by obtaining 
the resonance condition as a function of 
wavelength for a dielectric sphere, and follows the 
formalism described in~\cite{Johnson:93,Teraoka:06a}. 
Starting from a very general construction, a continuous electric field, oscillating with frequency $\omega$, takes the form 
$\mathbf{E}(\mathbf{r}) = 
\mathbf{E}_0(\mathbf{r})\,\mathrm{e}^{i \omega t}$, for position 
$\mathbf{r}$. This satisfies the wave equation: 
\begin{equation}
\mathbf{\mathbf{\nabla}}\times\mathbf{\mathbf{\nabla}}
\times\mathbf{E}_0(\mathbf{r}) 
- k^2 n^2(\mathbf{r}) 
\mathbf{E}_0(\mathbf{r}) = 0,  
\end{equation}
for a refractive index $n(\mathbf{r})$, and 
wave-number $k=\omega/c$, where $c$ is the speed of light in a vacuum. 
The TE modes are obtained by separation of variables in spherical polar coordinates 
\begin{align}
\mathbf{E}_0 &= 
S_m(r) \frac{\mathrm{e}^{i m \varphi}}{k\,r}\,\mathbf{X}_{lm}(\theta),\\
\mathbf{X}_{lm}(\theta) &=\frac{i\,m}{\sin\theta}
P_l^m(\cos\theta)\hat{\mathbf{e}}_\theta - 
\frac{\partial}{\partial\theta}P_l^m(\cos\theta)\hat{\mathbf{e}}_\varphi, 
\end{align}
where the angular vector function $\mathbf{X}_{lm}(\theta)$ is 
defined in terms of the associated Legendre polynomial $P_l^m$, and contains 
contributions in directions $\hat{\mathbf{e}}_\theta$ and 
$\hat{\mathbf{e}}_\varphi$. 
The function $S_m(r)$ satisfies the second order differential equation
\begin{equation}
\frac{\mathrm{d}S_m(r)}{\mathrm{d} r^2} + \Big(k^2\,n^2(r) - \frac{m(m+1)}{r^2}\Big)S_m(r)=0. 
\end{equation}

Up to this point, the electric field simply takes the general form of a field in spherical 
polar coordinates, embedded in a dielectric medium. Now, the resonance condition may be obtained 
for a dielectric sphere, by specifying continuity across the boundary 
in the radial component of the electric field, $S_m(r)$. 
For a refractive index of $n_1$ inside a sphere of radius $R$ 
($r\equiv|\mathbf{r}|<R$), and $n_2$ outside, 
the radial function $S_m(r)$ takes a Riccati-Bessel form, 
which can be expressed in terms of the spherical Bessel functions 
of the first kind, $j_m$, and the spherical Hankel function, $h_m^{(1)}$: 
\begin{equation}
S_m(r) = \Big\{
\begin{array}{ll} 
 z^r_1 \,\,j_m(z^r_1), \qquad & r< R \\
 A_m \, z^r_2 \,\,h_m^{(1)}(z^r_2) 
,\qquad & r> R. 
\end{array}
\end{equation}
The use of the outgoing spherical Hankel function, $h_m^{(1)}(z)$ 
 takes into account leaky WGMs, which radiate energy outwards, and are 
described by quasi-modes \cite{PhysRevA.41.5187} that are normalizable. 
The arguments $z^r_{1,2} \equiv  n_{1,2} \,k\, r$ are the so-called 
size parameters, and
 $A_m$ are the coefficients defined through continuity at the 
boundary. 
The resonance condition for the TE modes of a sphere of 
radius $R$ is thus \cite{Teraoka:06a} 
\begin{equation}
 \frac{n_1}{z^R_1} \frac{(m+1) j_m(z^R_1) - z^R_1\, j_{m+1}(z^R_1)}{j_m(z^R_1)} = 
\frac{n_2}{z^R_2} \frac{(m+1) y_m(z^R_2) -  z^R_2\, y_{m+1}(z^R_2)}{y_m(z^R_2)}.
\end{equation}

For TM modes, the form of $\mathbf{E}_0$ contains both angular and radial 
vector components (for $\hat{\mathbf{e}}_r$ as the unit vector in the radial direction)
\begin{align}
\mathbf{E}_0 &= \frac{\mathrm{e}^{i m \varphi}}{k^2\,n^2(r)}\left[
\frac{1}{r}\frac{\partial T_m(r) }{\partial r}
\mathbf{Y}_{lm}(\theta) + \frac{1}{r^2}T_m(r)\mathbf{Z}_{lm}(\theta)\right], \\
\mathbf{Y}_{lm}(\theta) &= \hat{\mathbf{e}}_r\times\mathbf{X}_{lm}(\theta),\\
\mathbf{Z}_{lm}(\theta) &= l(l+1)P_l^m(\cos\theta)\hat{\mathbf{e}}_r, 
\end{align}
where the function $T_m(r)$ obeys the equation
\begin{equation}
\label{eqn:Teqn}
\frac{\mathrm{d}T_m(r)}{\mathrm{d} r^2}
-\frac{2}{n(r)}\frac{\mathrm{d} n(r)}{\mathrm{d}r}\frac{\mathrm{d}T_m(r)}{\mathrm{d} r} 
+ \Big(k^2\,n^2(r) - \frac{m(m+1)}{r^2}\Big)T_m(r)=0. 
\end{equation}
The solutions to Eq.~(\ref{eqn:Teqn}) for $T_m(r)$ take a similar form to those of $S_m(r)$ above
\begin{equation}
T_m(r) = \Big\{
\begin{array}{ll} 
 z^r_1 \,\,j_m(z^r_1), \qquad & r< R \\
 B_m \, z^r_2 \,\,h_m^{(1)}(z^r_2) 
,\qquad & r> R, 
\end{array}
\end{equation}
where the coefficient $B_m$, analogously to $A_m$, may be defined through enforcing continuity at the boundary. 
This leads to the resonance condition 
\begin{equation}
 \frac{1}{n_1 z^R_1} \frac{(m+1) j_m(z^R_1) - z_1\, j_{m+1}(z^R_1)}{j_m(z^R_1)} = 
\frac{1}{n_2 z^R_2} \frac{(m+1) y_m(z^R_2) -  z_2\, y_{m+1}(z^R_2)}{y_m(z^R_2)}.
\end{equation}
By solving for the wave number $k$, a spectrum of independent TE and TM 
modes can be obtained, and compared to the FDTD simulation.

A more general type of analytic model, such as that 
developed by Chance, Prock and Silbey 
\cite{chance1978molecular} and separately by Chew \cite{PhysRevA.13.396,Chew:1987a,PhysRevA.38.3410}, 
is able to simulate the full electric and magnetic field behavior both inside and outside the resonator. 
This type of model will serve as an excellent benchmark comparison for the FDTD profile spectrum. 
Both models were shown to be equivalent formulations of spherical resonators excited by a 
dipole source \cite{Chew:1987a}, 
and obtain the same resonance positions as Mie scattering \cite{BohrenHuffman}. 
The simpler analytic models, however, are still useful at identifying the mode numbers of the peaks 
corresponding to WGMs in a microsphere. 
For a tangentially ($E_y$) or radially ($E_x$) oriented dipole source 
on the surface of a sphere of radius $R$ and refractive 
index $n_{1}=\sqrt{\epsilon_{1}\mu_{1}}$, 
the power output equations take the following form (when
 normalized to the surrounding medium of index $n_2$): 
\begin{align}
\label{eq:chewey}
P^{E_y}/P^{E_y}_0 &= \frac{3 \epsilon^{3/2}_1 n_1}{2\, {z^R_1}^2}\left(
\frac{\epsilon_2}{\mu_2}\right)^{1/2}\sum_{m=1}^\infty m(m+1)(2m+1)\frac{j_m\,^2(z^R_1)}{{z^R_1}^2\,|D_m|^2}, \\
\label{eq:chewex}
P^{E_x}/P^{E_x}_0 &= \frac{3 \epsilon^{3/2}_1 n_1}{4\, {z^R_1}^2}\left(
\frac{\epsilon_2}{\mu_2}\right)^{1/2}\sum_{m=1}^\infty(2m+1)\left[
\Big|\frac{[z^R_1\,\,j_m(z^R_1)]'}{z^R_1\,D_m}\Big|^2 
+ \frac{\mu_1 \mu_2}{\epsilon_1 \epsilon_2}\frac{j_m\,^2(z^R_1)}{|\tilde{D}_m|^2}\right], \\
\mbox{for}\quad D_m &= \epsilon_1 j_m(z^R_1)[z^R_2\,h^{(1)}_m(z^R_2)]' - 
\epsilon_2 h^{(1)}_m(z^R_2)[z^R_1\,j_m(z^R_1)]',\\
\tilde{D}_m &= D_m(\epsilon_{1,2}\rightarrow\mu_{1,2}). 
\end{align}
Although this analytic model is better 
suited to simulating scenarios  
that involve dipole sources at a variety of positions and alignments, 
the FDTD method is able to collect flux from any desired collection 
region (in the 
near or far field)
for any length of time.  
 This is an important point, 
as this collection scenario will affect the measured  
power spectrum profile. It is useful to be able to estimate the size of 
the effect due to collection region, and this will be investigated 
in the specific case of the angular distribution of the flux, 
in Section~\ref{sec:results}.

\section{Results}
\label{sec:results}

An FDTD simulation of a polystyrene ($n_1 = 1.59$) microsphere, 
with a diameter of $6$ $\mu$m is carried out for a tangentially-oriented 
electric dipole source, emitting a Gaussian pulse 
with a central wavelength of $600$ nm, and a width of $5$ fs. Note that the pulse width is 
significantly narrower than the decay transition rate expected from a typical fluorescent source, such as Rhodamine 
dye, which is roughly $1$--$3$ ns \cite{Barnes:92,pmid19950338}. In this work, the pulse is taken to be effectively 
instantaneous with respect to the phenomena being measured.   
A circular flux collection region with a diameter of $2.58$ $\mu$m is 
placed a distance of $240$ nm from the surface of the sphere in the 
$x$-direction, with its normal aligned along the same axis.  
Spectral information is then collected for wavelengths in the range $500$-$750$ nm. The finite grid resolution entails an asymmetry in the 
Gaussian peak, which diminishes as the resolution increases. 

\begin{widetext} 

\begin{figure}[tp]
\hspace{-5.5mm}
\includegraphics[width=0.503\hsize,angle=0]{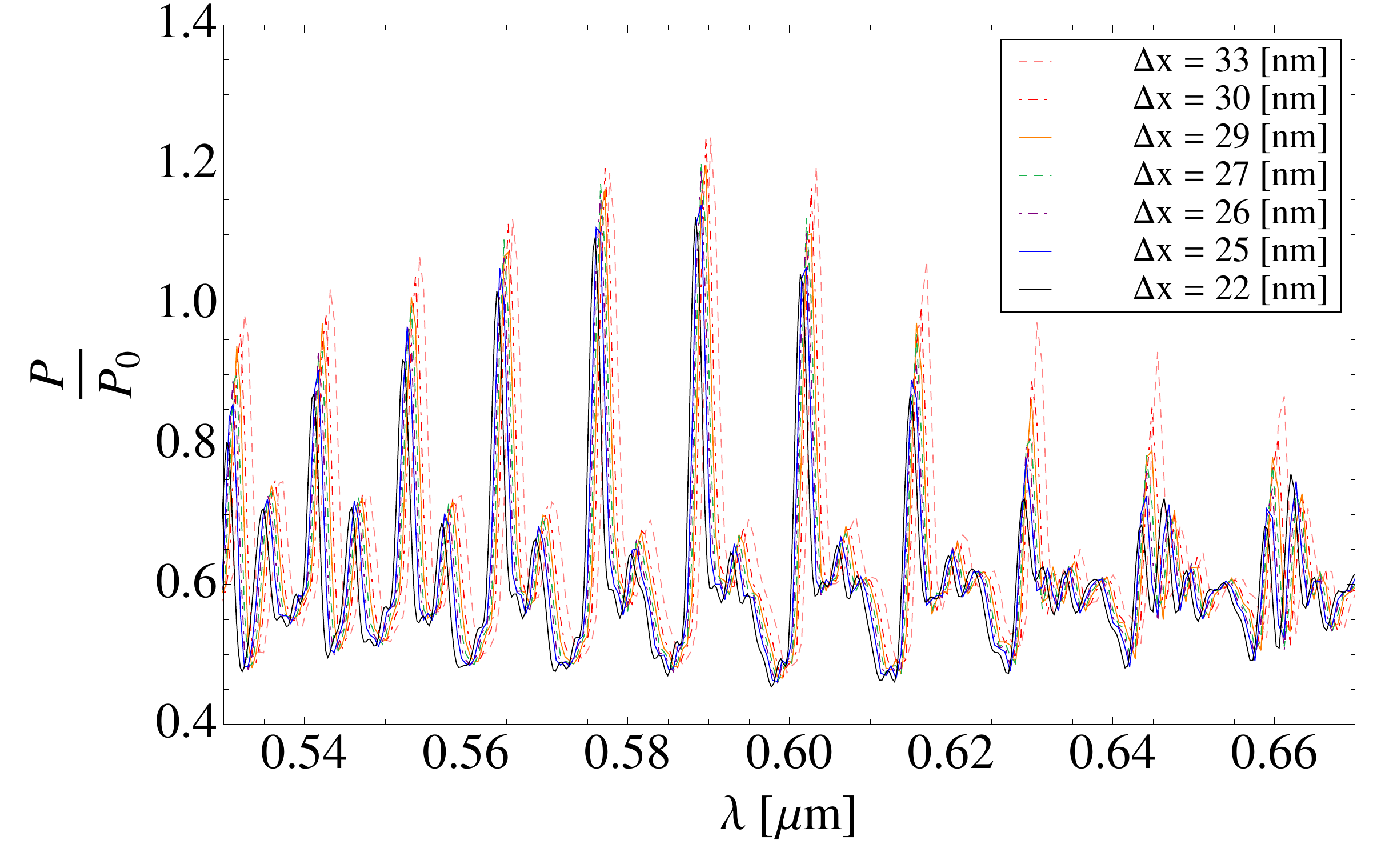}
\vspace{-2mm} 
\includegraphics[width=0.503\hsize,angle=0]{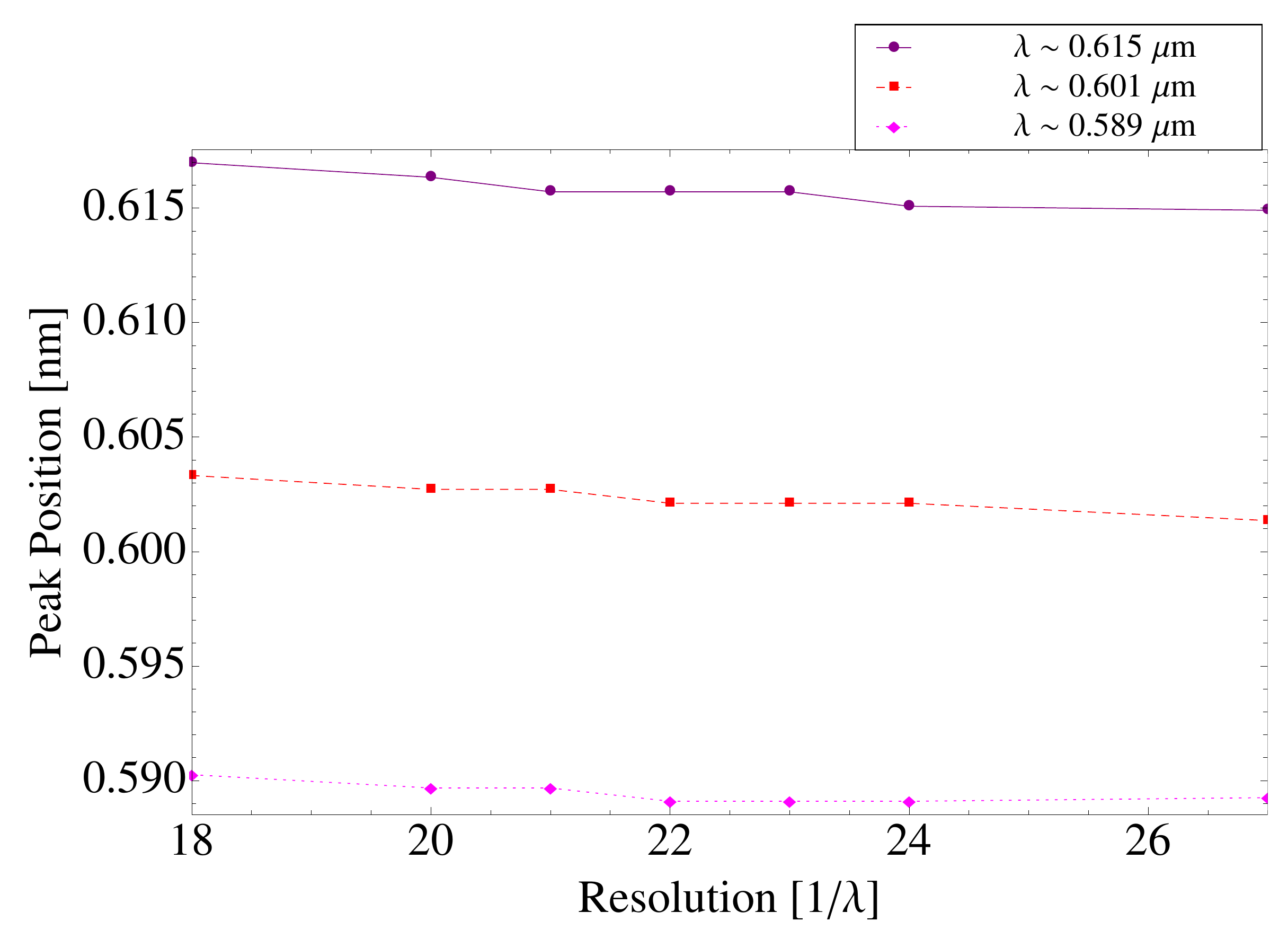}\\
\mbox{\hspace{3.1cm}(a)\hspace{6.3cm}(b)}
\vspace{-3mm}
\caption{\footnotesize{(a) A comparison of the power spectra of $6$ $\mu$m diameter microspheres for different grid resolutions. (b) The convergence of the position of three central peaks is shown as a function of resolution. Excitation occurs from a tangentially oriented dipole source placed on the surface, with a central wavelength of $\lambda = 0.6$ $\mu$m. The flux collection time for each simulation 
is $0.6$ ps.}} 
\label{fig:res}
\end{figure}
\end{widetext}

A comparison of a variety of grid resolutions is shown in  Fig.~\ref{fig:res}(a), 
using a total flux collection time of $0.6$ ps ($120$ times the width of the pulse). 
For resolutions in each spatial direction of $22-33$ nm, 
one finds that as the resolution decreases, the profile features of the spectrum do not 
alter significantly; there is only a small offset in the positions of the 
peaks, 
and the peak heights. The positions of the WGMs are determined from a flux collection 
of the frequencies, which are then converted to wavelengths. The temporal resolution 
is further interpolated to yield a value $< 1$ nm. 

The systematic uncertainty in the resonance positions due to resolution 
can be quantified by tracking the positions of the WGM peaks. 
By examining the wavelength positions of the 
most prominent peaks, 
one can obtain a converged result to a chosen tolerance, 
as shown in Fig.~\ref{fig:res}(b).  
The convergence of the peak positions can be obtained 
by observing a systematic shift in the positions as the resolution is improved. 
%NEW
For example, using a resolution of 
$\Delta x = 25$ nm, the most central peak has a wavelength position of 
$602.11$ nm. By improving the resolution to $\Delta x = 22$ nm, the position 
becomes $601.35$ nm.  This yields an 
improvement in the position of the central 
peak of $(602.11-601.35) = 0.76$ nm, or $0.13\%$. 
This indicates a reasonable level of convergence 
of the peaks, as any further increase in resolution is expected to improve 
the result by $0.13\%$ at most. 
Note that an even greater spectral resolution ($\sim~0.03$~nm) 
is required 
for the detection of 
nearby macromolecules by a microsphere 
\cite{PMID:22219711}. This presents an issue in 
 simulating the effect of detection using FDTD. However, 
 FDTD is suitable for providing 
a realistic estimate of the WGM $Q$-factors under a variety of 
 resonator scenarios, and for assessing their feasibility for 
new designs of biosensors.

\begin{widetext} 

\begin{figure}[tp]
\hspace{-5.5mm}
\includegraphics[width=0.503\hsize,angle=0]{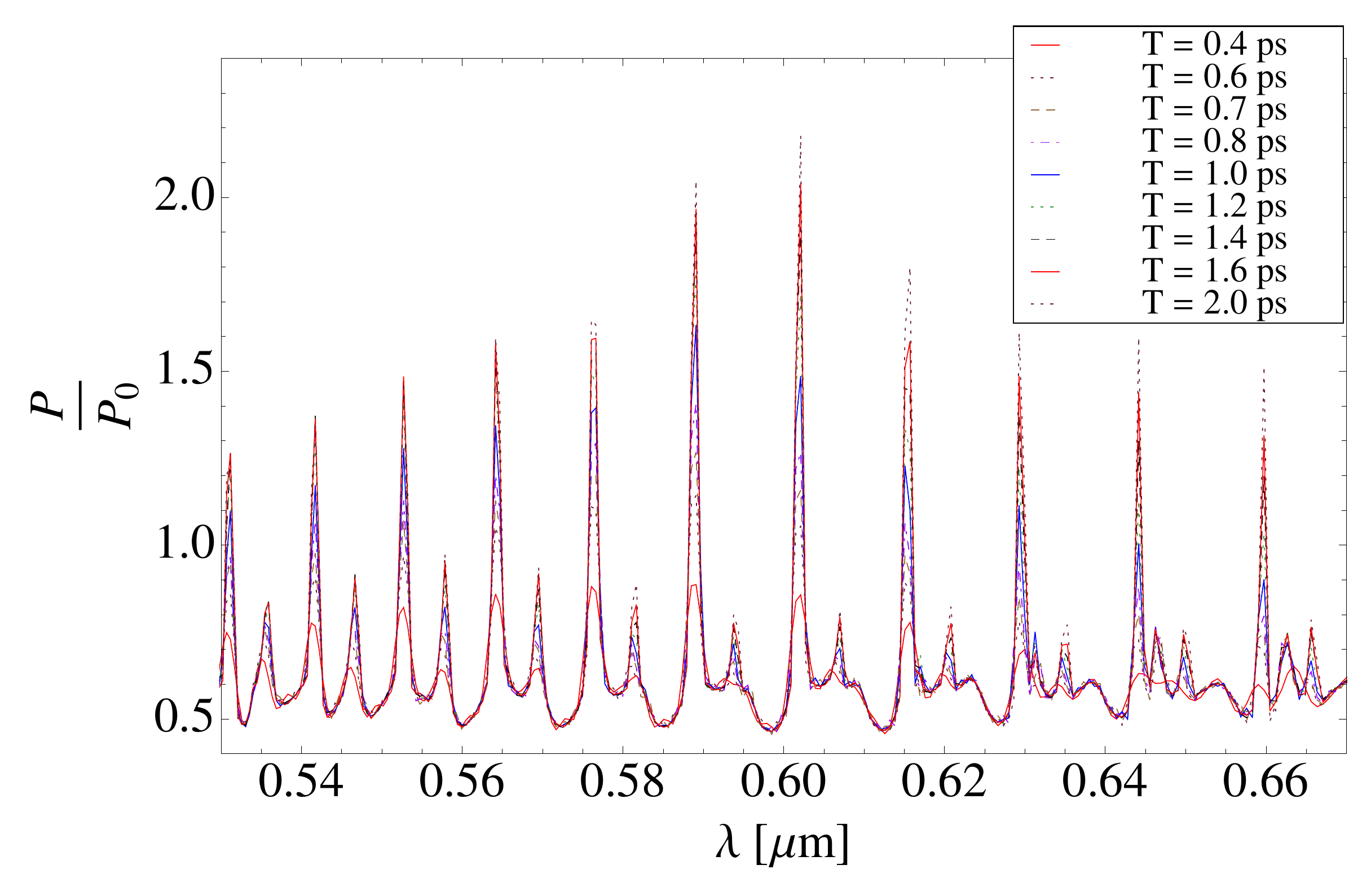}
\vspace{-2mm}
\hspace{1mm}
\includegraphics[width=0.503\hsize,angle=0]{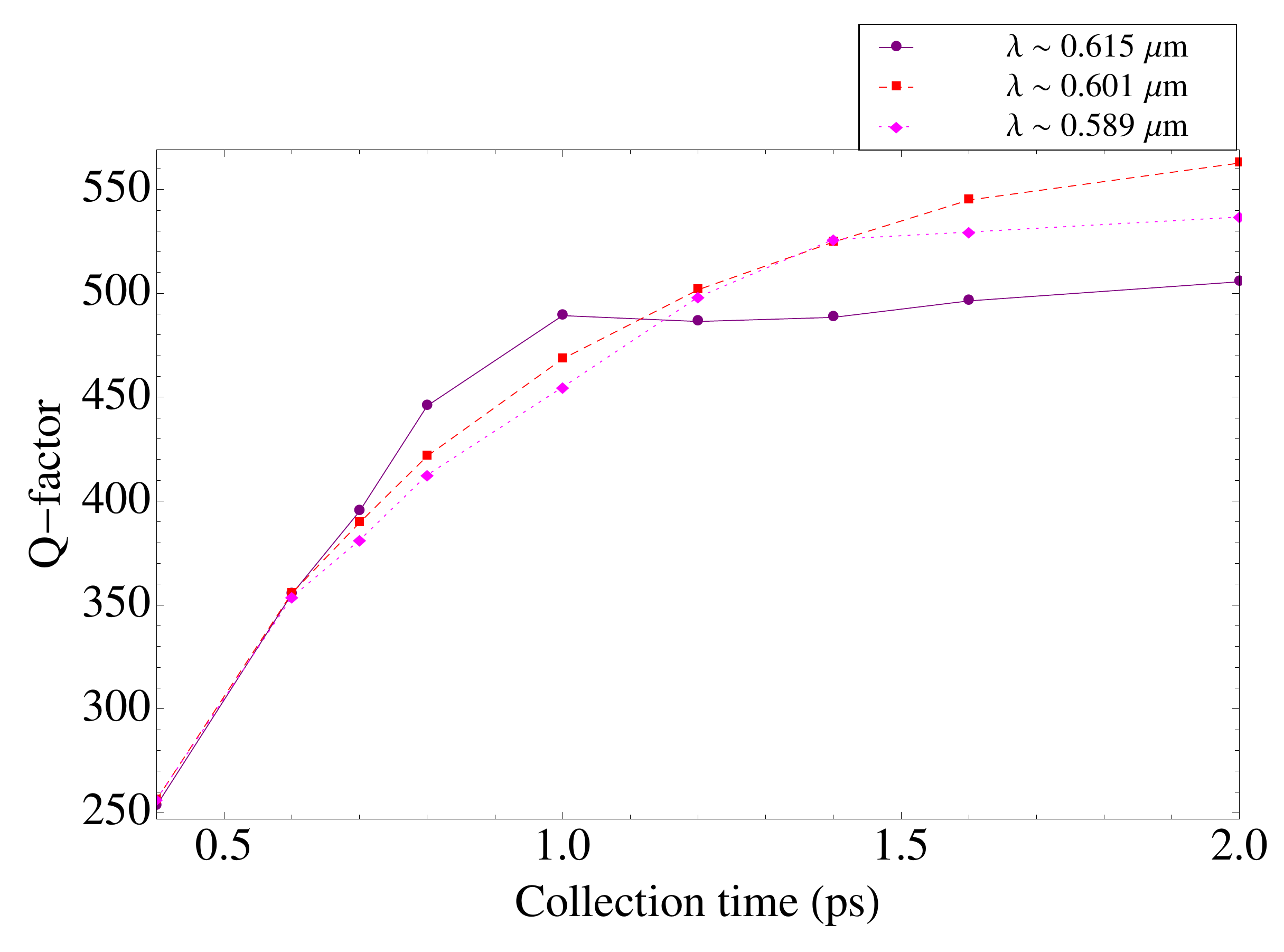}
\\
\mbox{\hspace{3.1cm}(a)\hspace{6.3cm}(b)}
\vspace{-3mm}
\caption{\footnotesize{(a) A comparison of the power spectra of $6$ $\mu$m diameter microspheres with a tangential source ($\lambda = 0.6$ $\mu$m), for different flux collection times. (b)  The behavior of the $Q$-factors of three central peaks is shown as a function of collection time (ps).  The output power is normalized to the dipole emission rate in an infinite bulk medium of the same refractive index as the surrounding medium. }}
\label{fig:WGMtime}
\end{figure}
\end{widetext}

By changing the
 length of time allowed for flux collection, one can also obtain important 
insights into the distribution of modes in the power spectrum. 
Figure~\ref{fig:WGMtime}(a) 
 shows the resultant normalized power spectrum for a variety of flux collection times.  
A spatial resolution of 
$\Delta x=25$ nm is chosen, and also  
a density of sampled wavelengths corresponding to 
$\Delta \lambda = 0.62$ nm. 
As the collection time increases, the WGM peaks become more prominent. 
Furthermore, 
double-peak structures seen at small collection times are no longer apparent for larger 
collection times, as the sampling of the full mode structure of the radiating 
cavity does not have sufficient 
time to be measured by the flux region. 
For the longer collection times, 
 an enhancement in the normalized dipole power is observed ($P/P_0 > 1$), 
at wavelengths coinciding with the 
dominant WGMs. This indicates that the power output 
of the source is enhanced at these wavelengths by the presence of the 
dielectric microsphere. 

The behavior of the $Q$-factor as a function of collection time 
is shown in Fig.~\ref{fig:WGMtime}(b) for three central 
WGM peaks. For the peak at $\lambda \sim 0.615$ $\mu$m, convergence is achieved 
beyond a collection time of $1.0$ ps, with a variation of $3.4\%$. 
For the other two peaks, $\lambda \sim 0.601$ $\mu$m and $0.589$ 
$\mu$m, variations of $16.9\%$ and $15.3\%$ are measured, respectively. 
Note that, while peaks with lower $Q$-factors incur less variation 
with respect to collection time, peaks with higher $Q$-factors are more 
affected. However, an increase in collection time becomes 
increasingly computationally intensive, due to the asymptotic behavior 
in the $Q$-factor 
with respect to collection time \cite{marki2007design}. Nevertheless, 
the uncertainty in the $Q$-factor due to the collection time can be 
quantified, and thus incorporated into the systematic uncertainty. 
This systematic effect will be considered when reporting the results
in Table~\ref{tab:qf}. 

\begin{widetext}

\begin{figure}[t]
\hspace{-5.5mm}
\includegraphics[width=0.503\hsize]{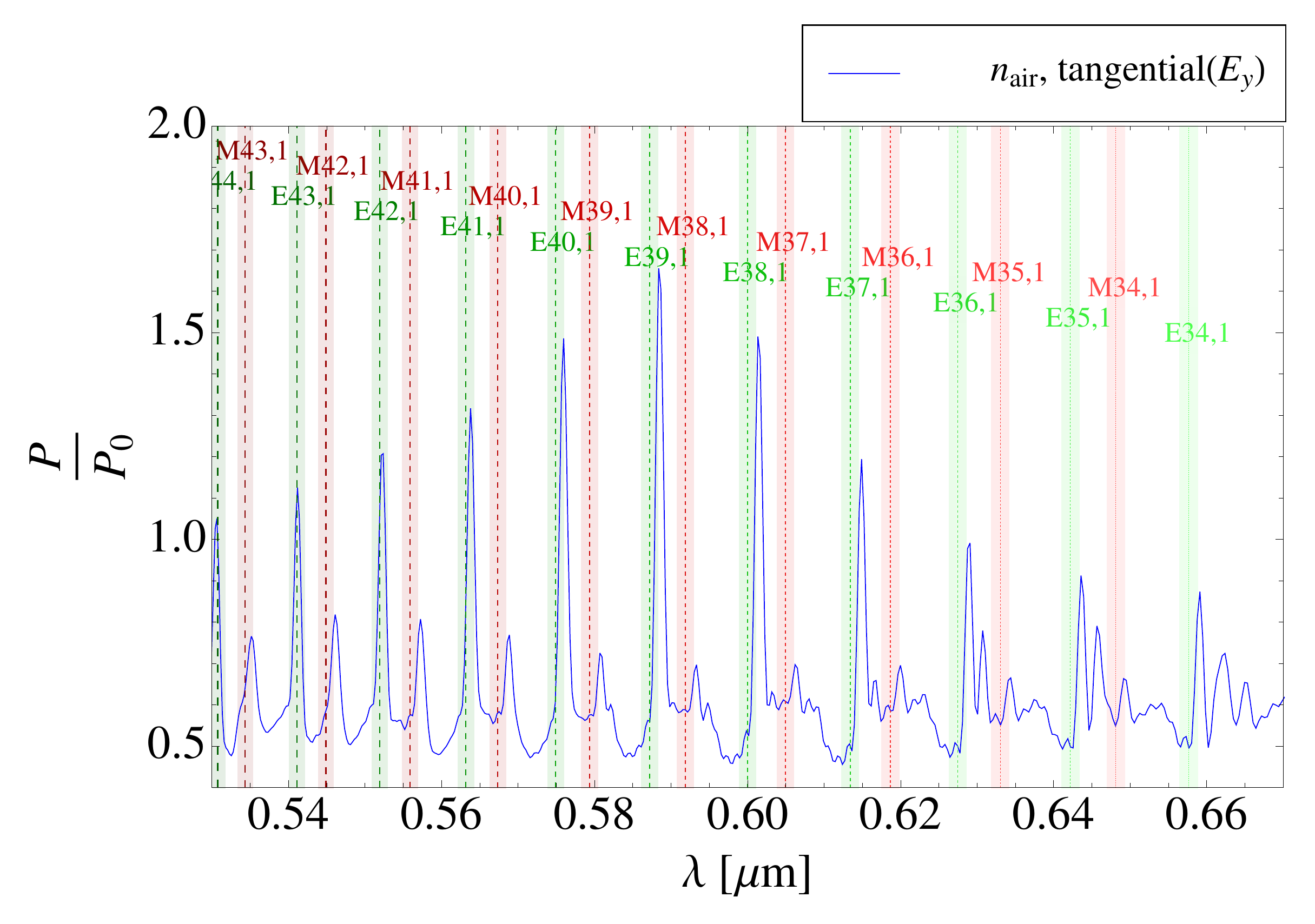}
\hspace{-3mm}
\vspace{-2mm}
\includegraphics[width=0.503\hsize]{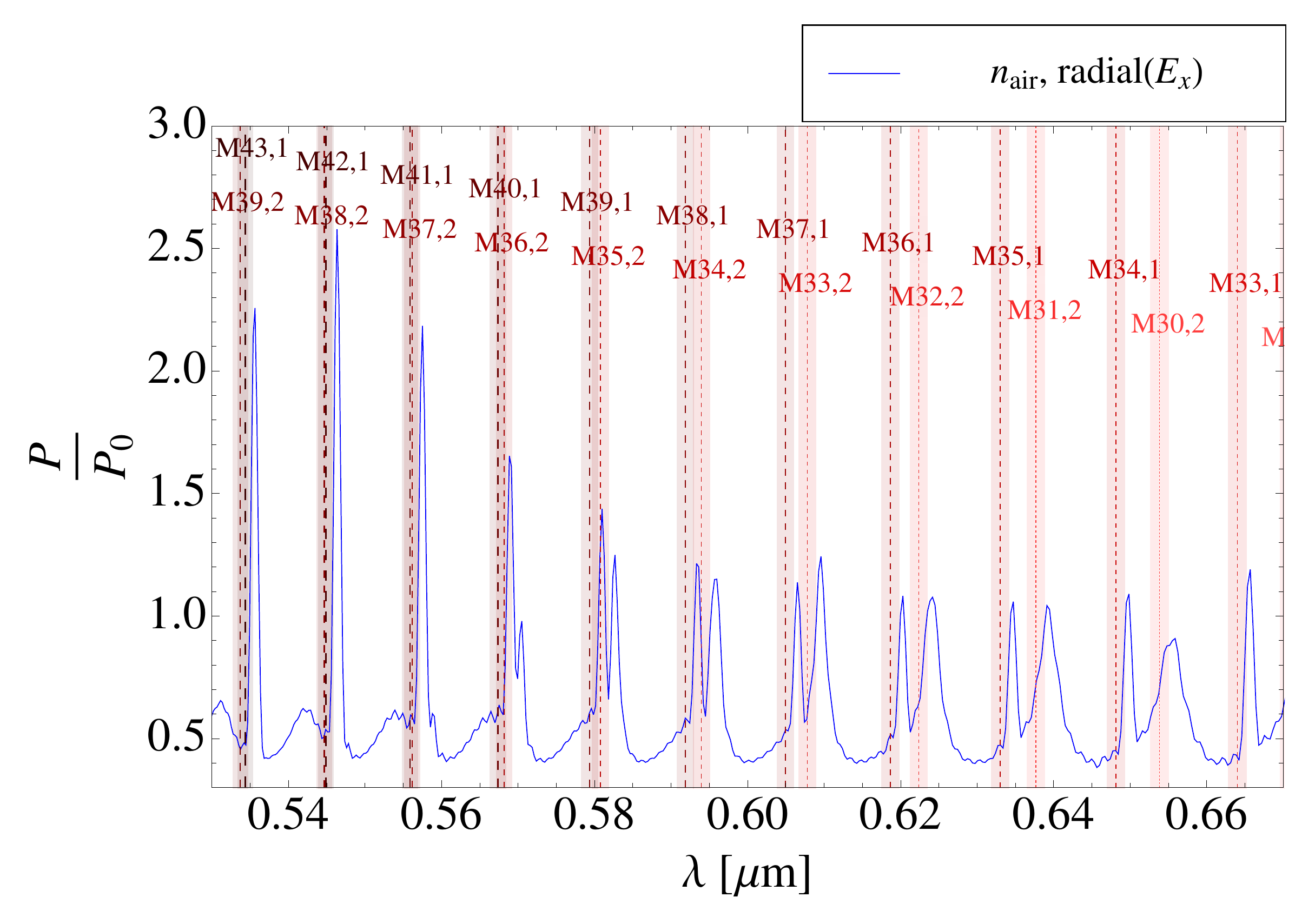}\\
\mbox{\hspace{3.1cm}(a)\hspace{6.2cm}(b)}\\
\vspace{1mm}
\hspace{-5.5mm}
\includegraphics[width=0.503\hsize]{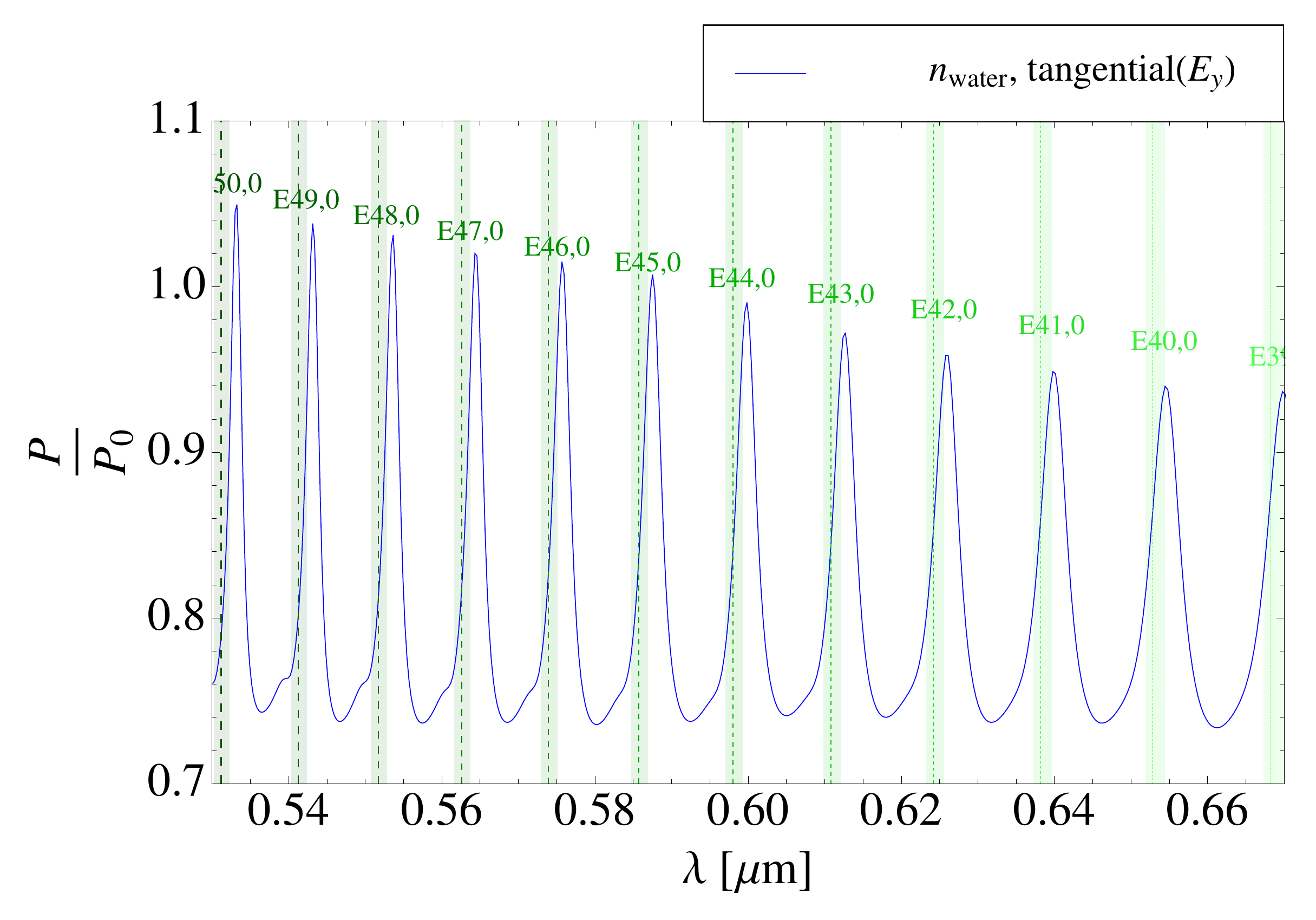}
\hspace{-3mm}
\vspace{-3mm}
\includegraphics[width=0.503\hsize]{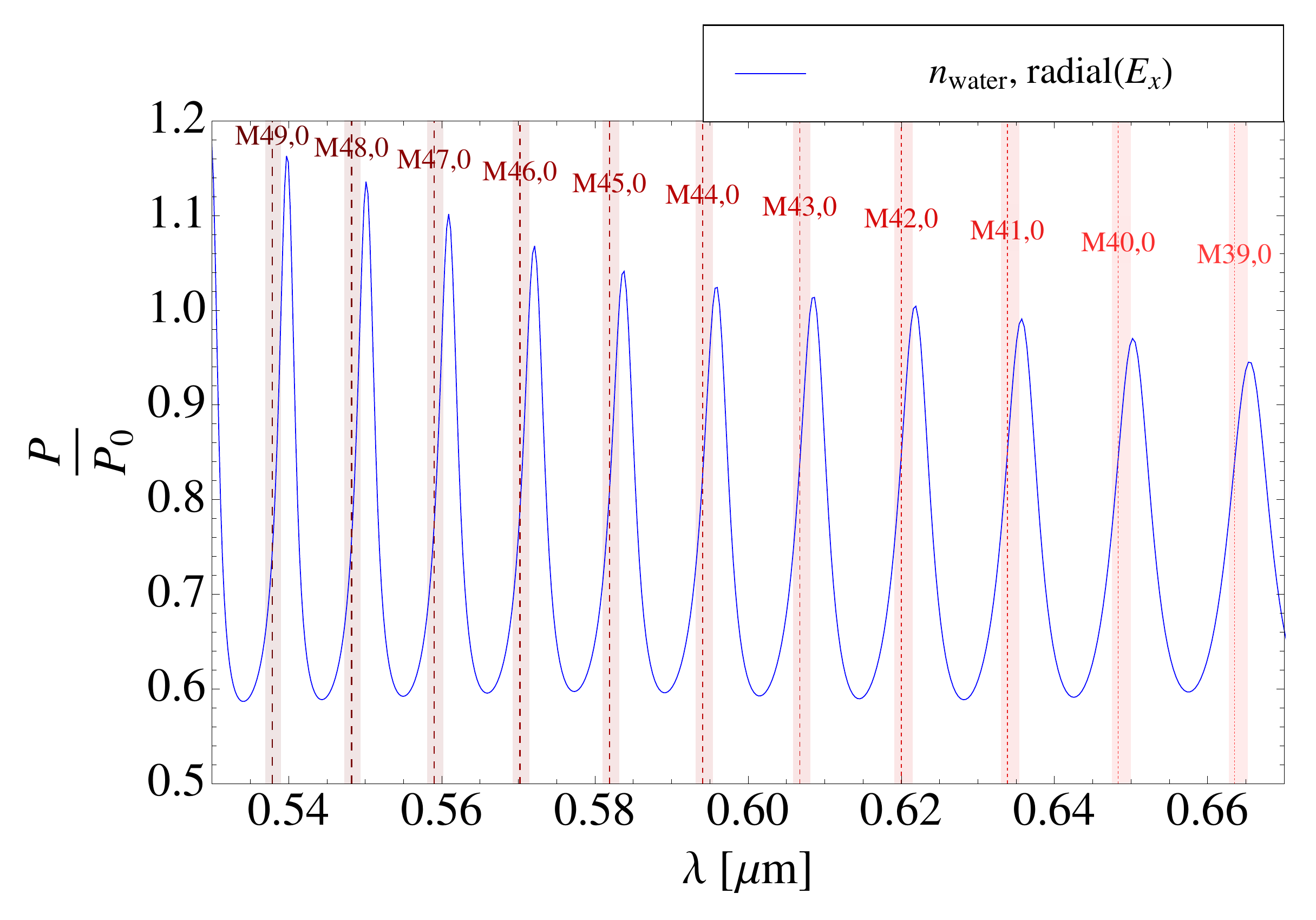}
\\
\vspace{-2mm}
\mbox{\hspace{3.1cm}(c)\hspace{6.2cm}(d)}\\
\vspace{-3mm}
\caption{\footnotesize{ FDTD simulation of the normalized power spectrum of a polystyrene microsphere, $6$ $\mu$m in diameter, with a surrounding medium of air. (a) Whispering gallery modes are excited from a tangential or (b) radial electric dipole source with a central wavelength of $0.6$ $\mu$m. 
(c)  The results for a surrounding medium of water are also shown for a 
tangential source and 
(d) a 
radial source. 
Vertical lines indicate predictions of the TE$_{m,n}$ (\textit{green}) and TM$_{m,n}$ (\textit{red}) modes derived from the 
Mie scattering analytic model, for azimuthal and radial mode numbers $m$ and $n$, respectively. The width of the bands indicates the systematic uncertainty in the positions due to the finite grid size of FDTD.}}
\label{fig:WGM}
\end{figure}
\end{widetext}

A comparison of the analytic 
model and the FDTD simulated spectrum for a microsphere 
is shown in Fig.~\ref{fig:WGM} for tangentially and radially 
oriented dipole sources in both air and water media. 
The WGM positions from the analytic model are shown 
as vertical bands, with TE$_{m,n}$ modes in green and TM$_{m,n}$ modes in red, 
for azimuthal and radial mode numbers $m$ and $n$, respectively. 
The width of each band indicates the estimate of the systematic uncertainty 
due to the finite grid spacing of FDTD. 
Specifically, the Mie scattering analytic model 
is used to estimate the shift in the WGM positions due to uncertainty in the 
sphere diameter, $6$~$\mu$m~$\pm\Delta x /2$. 
In this case, the spatial 
resolution is held fixed at the value $\Delta x=22$ nm. 
For the temporal resolution, 
the spectral density of $800$ points yields an uncertainty of 
$(750-500)$nm$/800 = 0.31$~nm. The two uncertainties are
 added in quadrature. 
The small offset of each peak from the position expected from 
the analytic model  may be due to mixing of nearby modes, which interact 
with each other to shift the peaks away from their central positions. 
The TE and TM modes cannot be completely 
decoupled due to the spherical symmetry, 
and contributions from both TE and TM modes are expected in the spectrum, 
in addition to non-whispering gallery radiation modes. 

\begin{table*}[h]
  \caption{\footnotesize{A summary of the $Q$-factors and wavelength positions, 
$\lambda$($\mu$m), 
 for the four most prominent WGM peaks, 
for each plot displayed in Fig.~\ref{fig:WGM}. 
The scenarios considered are: a surrounding medium of air or water   
with a tangential ($E_y$) or radial ($E_x$) dipole source. Due to 
finite collection time, a systematic uncertainty of up to $17\%$ is expected 
in the $Q$-factors. 
}}
\vspace{-6pt}
  \newcommand\T{\rule{0pt}{2.8ex}}
  \newcommand\B{\rule[-1.4ex]{0pt}{0pt}}
  \begin{center}
    \begin{tabular}{lllll}
      \hline
      \hline
      \T\B            
       Scenario &  peak 1: $\lambda$ ($\mu$m), $Q$  & peak 2: $\lambda$ ($\mu$m), $Q$  & peak 3: $\lambda$ ($\mu$m), $Q$  & peak 4: $\lambda$ ($\mu$m), $Q$  \\
      \hline     
      air, $E_y$ &\qquad\quad\,$0.576$, $490$ &\qquad\quad\,$0.588$, $510$ &\qquad\quad\,$0.601$, $528$ &\qquad\quad\,$0.615$, $559$\\
air, $E_x$ &\qquad\quad\,$0.536$, $549$  &\qquad\quad\,$0.546$, $572$ &\qquad\quad\,$0.557$, $584$ &\qquad\quad\,$0.569$, $584$\\
water, $E_y$ &\qquad\quad\,$0.533$, $305$ &\qquad\quad\,$0.543$, $286$ &\qquad\quad\,$0.553$, $264$ &\qquad\quad\,$0.564$, $250$ \\
      water, $E_x$ &\qquad\quad\,$0.539$, $232$ &\qquad\quad\,$0.550$, $218$ &\qquad\quad\,$0.560$, $204$ &\qquad\quad\,$0.572$, $191$  \\ 
      \hline
    \end{tabular}    
  \end{center}
  \label{tab:qf}
\end{table*} 
 
For a tangential electric dipole (Fig.~\ref{fig:WGM}(a)), 
one expects that the dominant WGMs excited are the 
lowest-order radial TE modes,  
since the electric field of each of the TM modes   
contains a radial component (see Section~\ref{sec:analytic}), 
which does not couple strongly to the tangential source. 
It is found that the dominant peaks have a free spectral range (FSR) 
consistent with a radial mode number of $n=1$. Note that the fundamental radial 
modes ($n=0$) cannot be resolved for this index contrast 
at this finite grid size,
since they are known to exhibit extremely large $Q$-factors experimentally 
\cite{Francois:2011} and in the Chew model \cite{PhysRevA.38.3410}. 
In the case of the radial electric dipole 
(Fig.~\ref{fig:WGM}(b)), 
the dominant peaks exhibit an FSR that matches the $n=1$ and $n=2$ TM
 modes. In this case, there appears to be very little contribution 
from the TE modes. 

For a lower index contrast scenario, such as polystyrene microspheres in 
a surrounding medium of water (Figs.~\ref{fig:WGM}(c) and \ref{fig:WGM}(d)), 
the WGM peaks are broadened, and there is a reduced density of modes 
at wavelengths in the range 
$500$-$750$ nm. As a result, the peaks observed in 
the 
FDTD simulation correspond to the fundamental radial TE/TM modes, for 
a tangential/radial dipole source, respectively.

\begin{widetext}

\begin{figure}[t]
\hspace{-5.5mm}
\includegraphics[width=0.495\hsize]{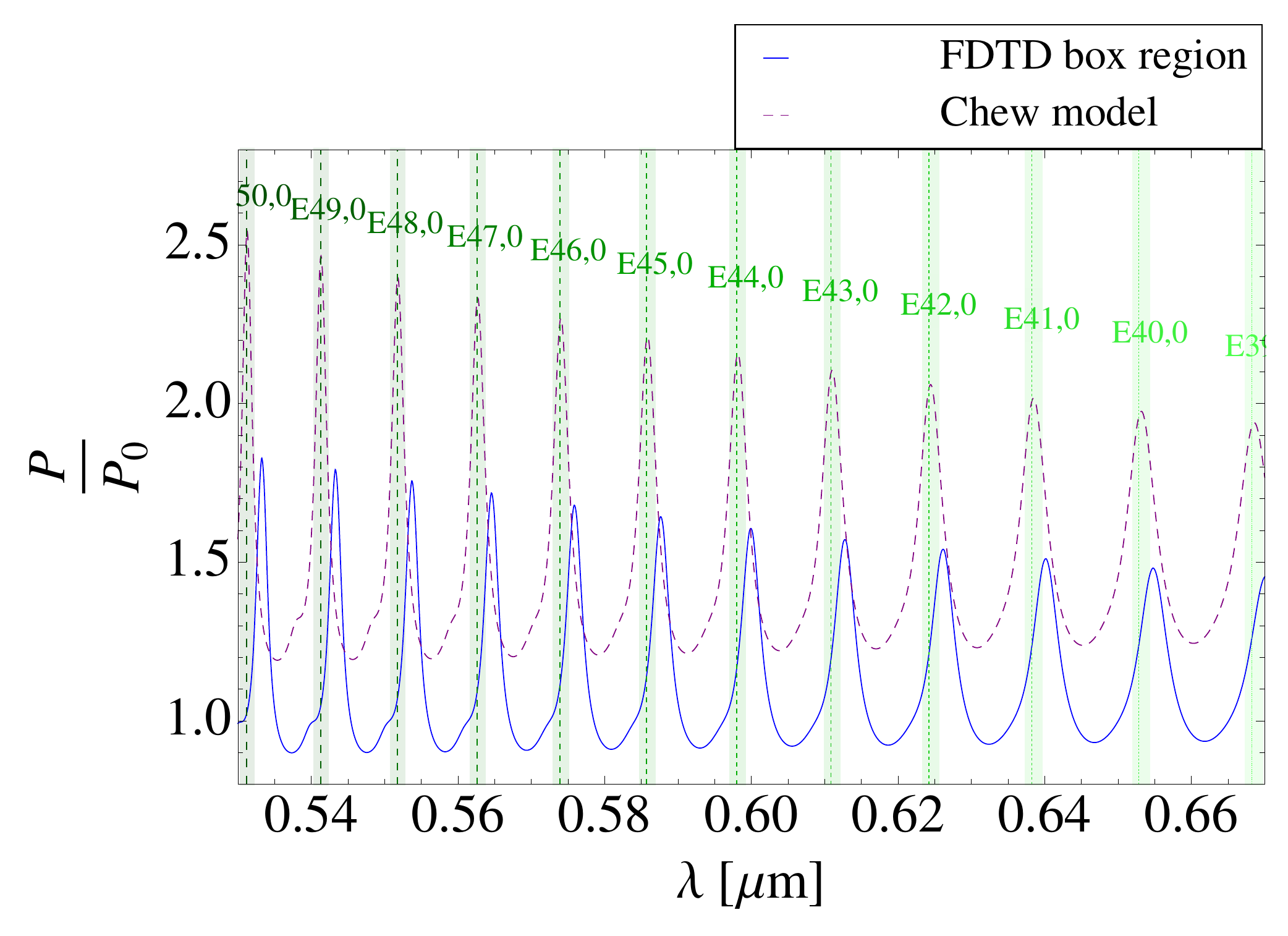}
\vspace{-2mm}
\includegraphics[width=0.495\hsize]{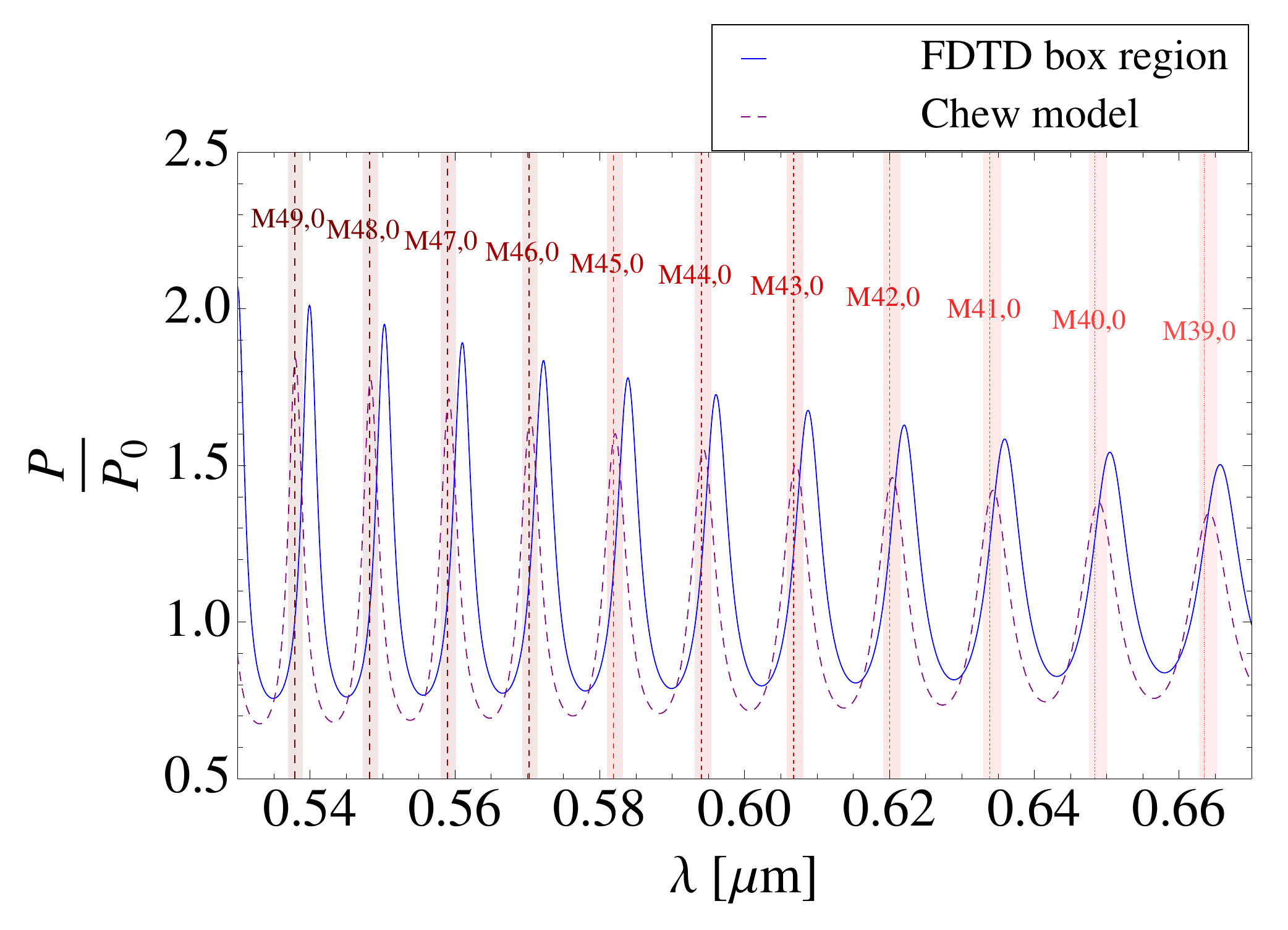}
\mbox{\hspace{3.35cm}(a)\hspace{6.5cm}(b)}
\vspace{-3mm}
\caption{\footnotesize{ 
A comparison of the Chew model of Eqs.~(\ref{eq:chewey}) \& (\ref{eq:chewex}) 
(dashed purple line) with the Mie scattering analytic model 
(vertical dashed lines) and the FDTD simulation, with total power collected 
from a box that surrounds the microsphere (solid blue line). 
A surrounding medium of water is used for a $6$ $\mu$m diameter polystyrene 
sphere. (a)  Whispering gallery modes are excited from a tangential or (b) a radial dipole source. 
The \textit{green} vertical lines are the fundamental radial TE modes, and the \textit{red} lines are the corresponding TM modes. The width of the bands accommodates the systematic uncertainty due to the finite grid size of FDTD.}}
\label{fig:Chew}
\end{figure}
\end{widetext}

Comparisons of the Chew model, the Mie scattering analytic model and 
the FDTD simulations are shown in Fig.~\ref{fig:Chew}, for a surrounding 
medium of water. In Fig.~\ref{fig:Chew}(a), 
a tangential dipole source is used, and the fundamental radial TE modes 
from the Mie scattering model (green vertical lines) 
exactly match the peaks of the Chew model (dashed blue line), 
as expected by construction \cite{PhysRevA.38.3410}. 
The peaks of the FDTD simulation also correspond to these TE modes, 
and the FSR matches that of the Chew model, with a small
 peak offset due to the finite grid resolution. 
Figure~\ref{fig:Chew}(b) shows the result for a radial 
dipole. The peaks of both the Chew model and the FDTD simulation 
correspond to the fundamental radial TM modes.

\begin{widetext}

\begin{figure}[!htbp]
\begin{center}
\includegraphics[width=0.4\hsize]{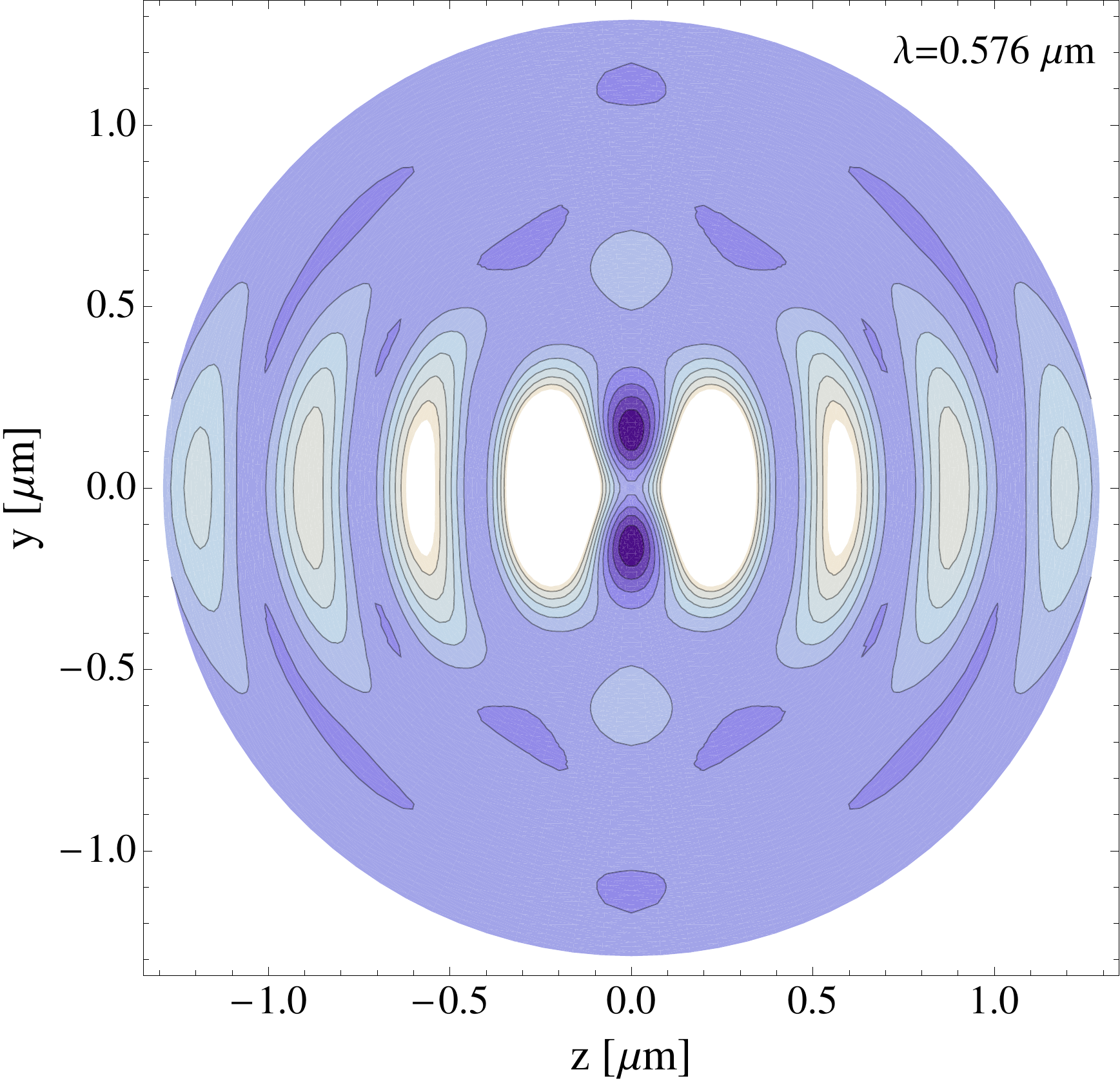}
\hspace{-1.0mm}
\includegraphics[width=0.4\hsize]{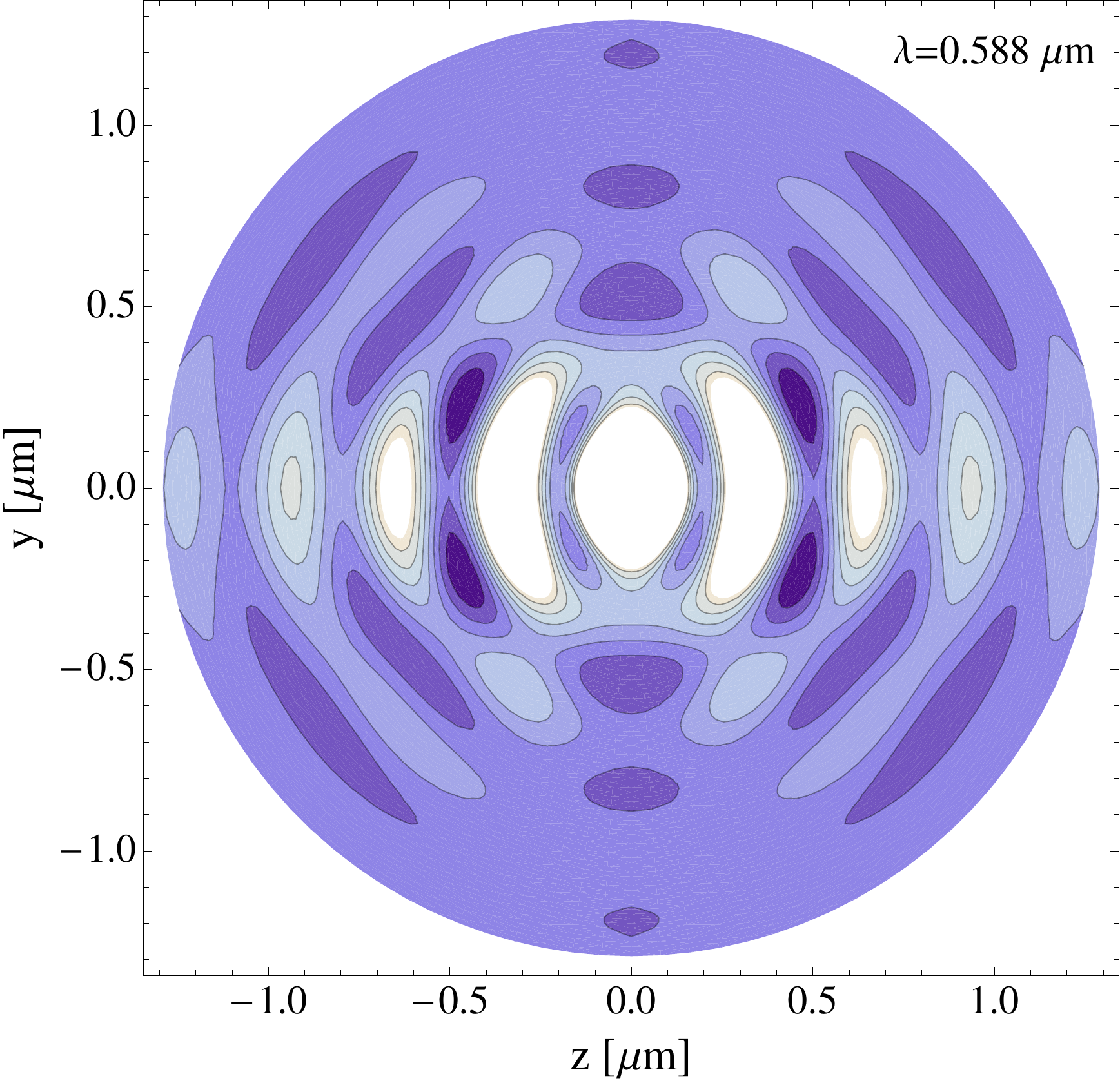} \\
\vspace{-1mm}
\mbox{\hspace{0.7cm}(a)\hspace{5.0cm}(b)}\\
\vspace{1mm}
\hspace{-1.0mm}
\includegraphics[width=0.4\hsize]{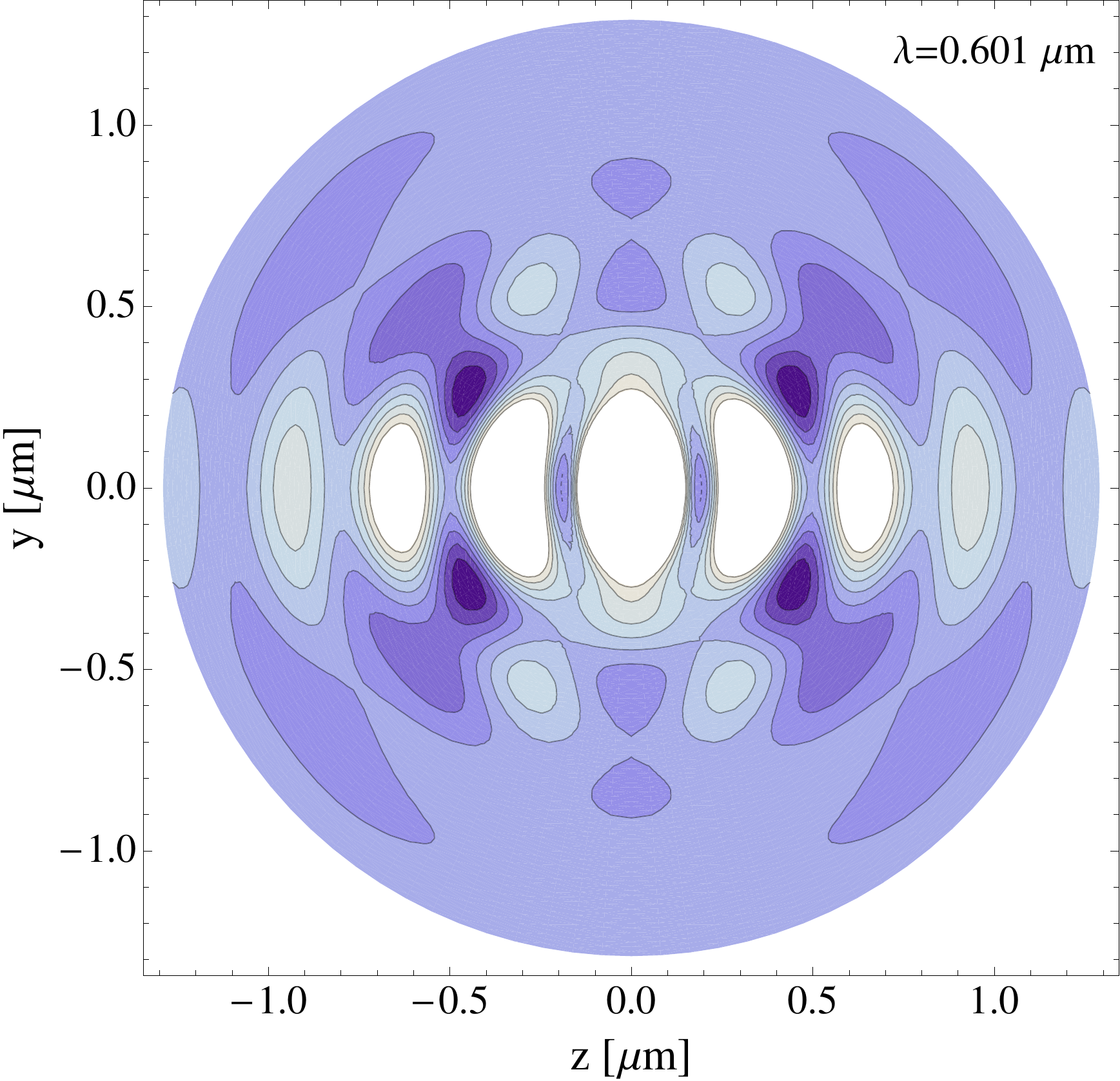}
\hspace{-1.0mm}
\includegraphics[width=0.4\hsize]{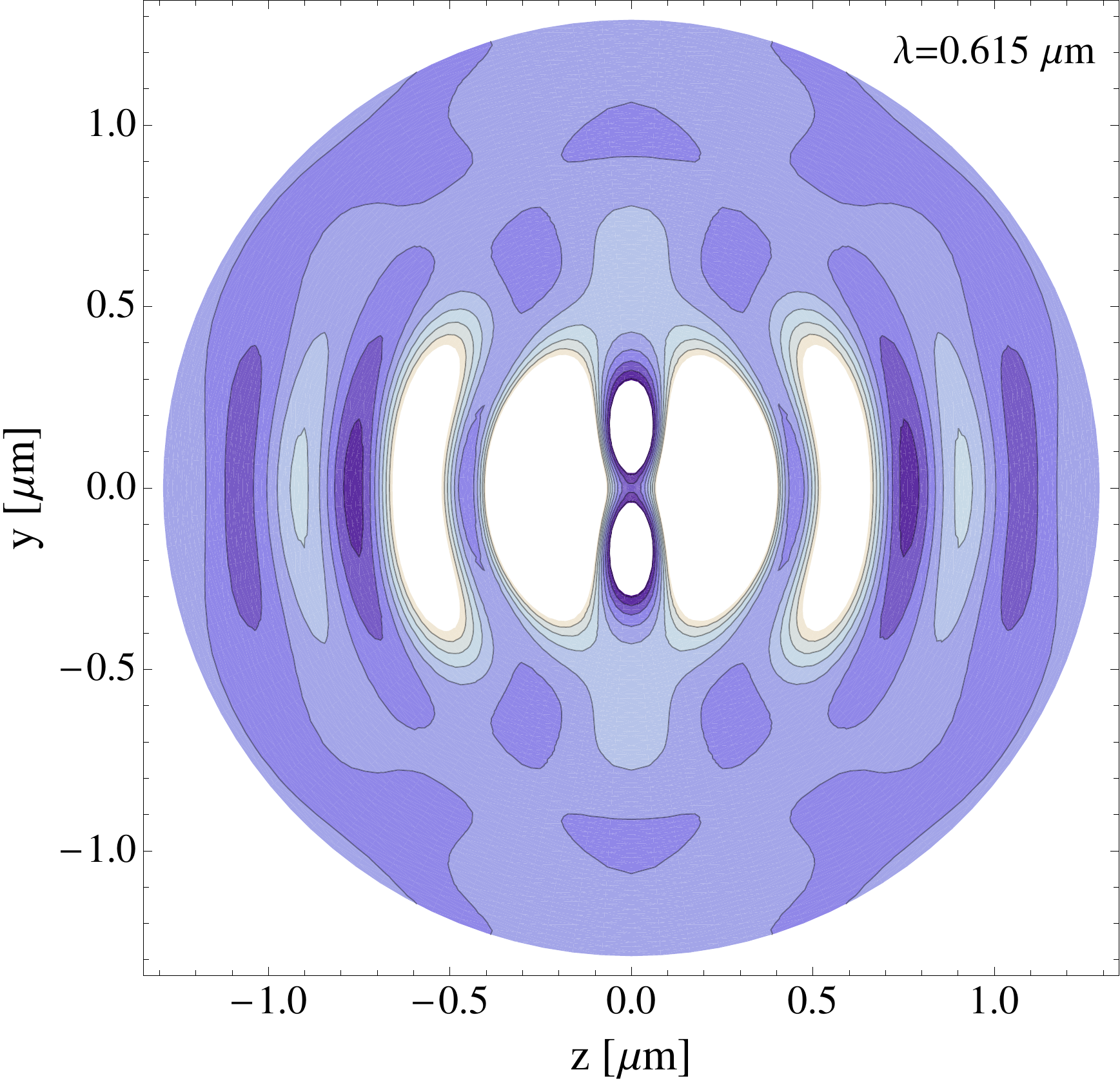}
\\
\vspace{-1mm}
\mbox{\hspace{0.7cm}(c)\hspace{5.0cm}(d)}
\vspace{-3mm}
\caption{\footnotesize{ Spatial distributions of the flux density over the collection region for several modes, 
$\mathbf{S}(\mathbf{r})$, integrated 
over a flux collection period of $1.0$ ps, and projected onto the circular flux collection region. 
The modes considered are (a)  TE$_{40,1}$: $0.576$ $\mu$m, (b) TE$_{39,1}$: $0.588$ $\mu$m, (c) 
TE$_{38,1}$: $0.601$ $\mu$m  and (d) TE$_{37,1}$: $0.615$ $\mu$m.  
The axes are defined in the same way as in Fig.~\ref{fig:Sphere}, for $x$-coordinate: $3.24$ $\mu$m. Lighter color corresponds to larger magnitudes of $\mathbf{S}(\mathbf{r})\cdot\hat{\mathbf{n}}$.}}
\label{fig:WGMrad}
\end{center}
%\end{figure}
%\end{widetext}
%
%\begin{widetext}
%
%\begin{figure}[b]
\hspace{-3.5mm}
\includegraphics[width=0.503\hsize]{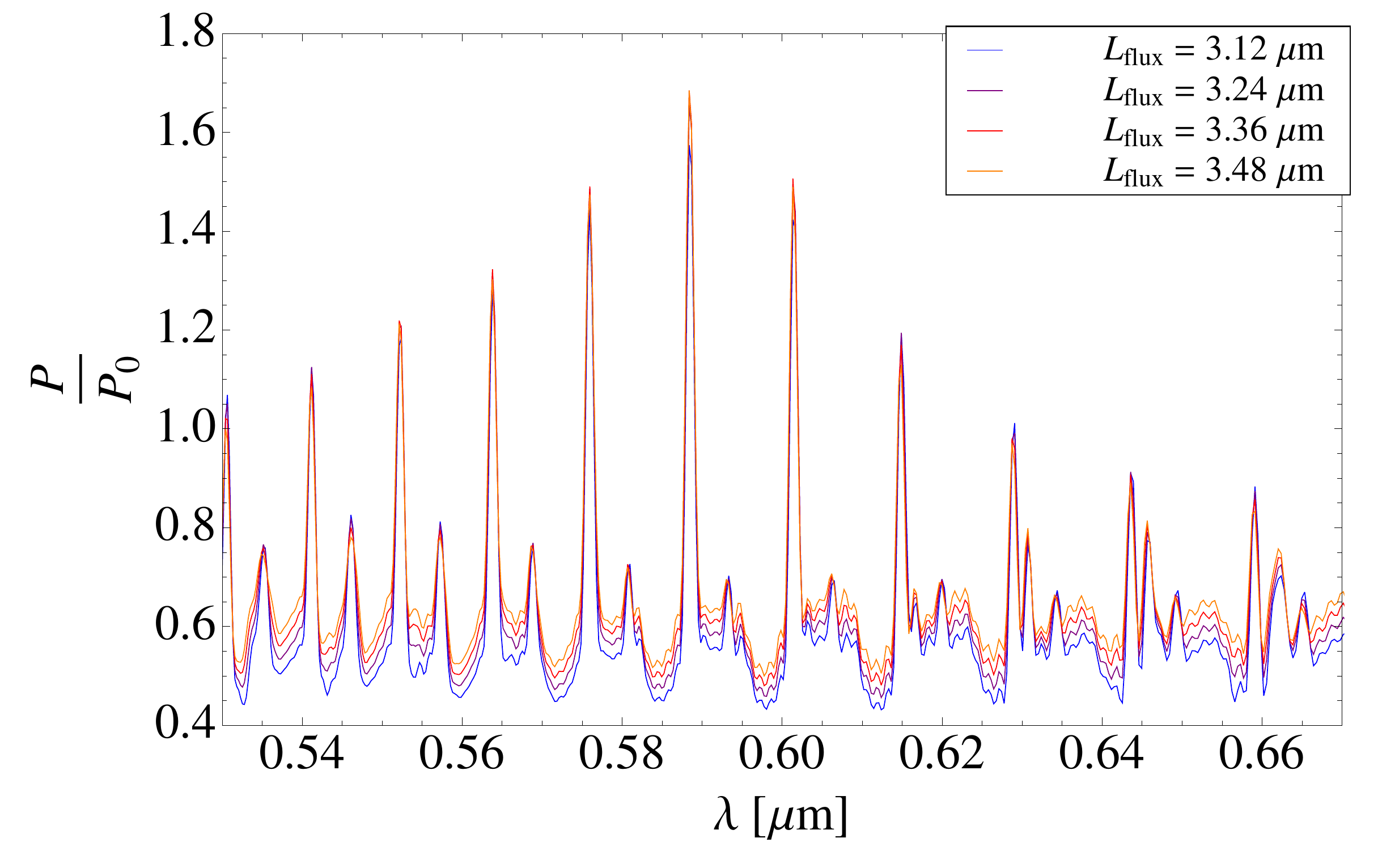}
\hspace{-3mm}
\vspace{-2mm}
\includegraphics[width=0.503\hsize]{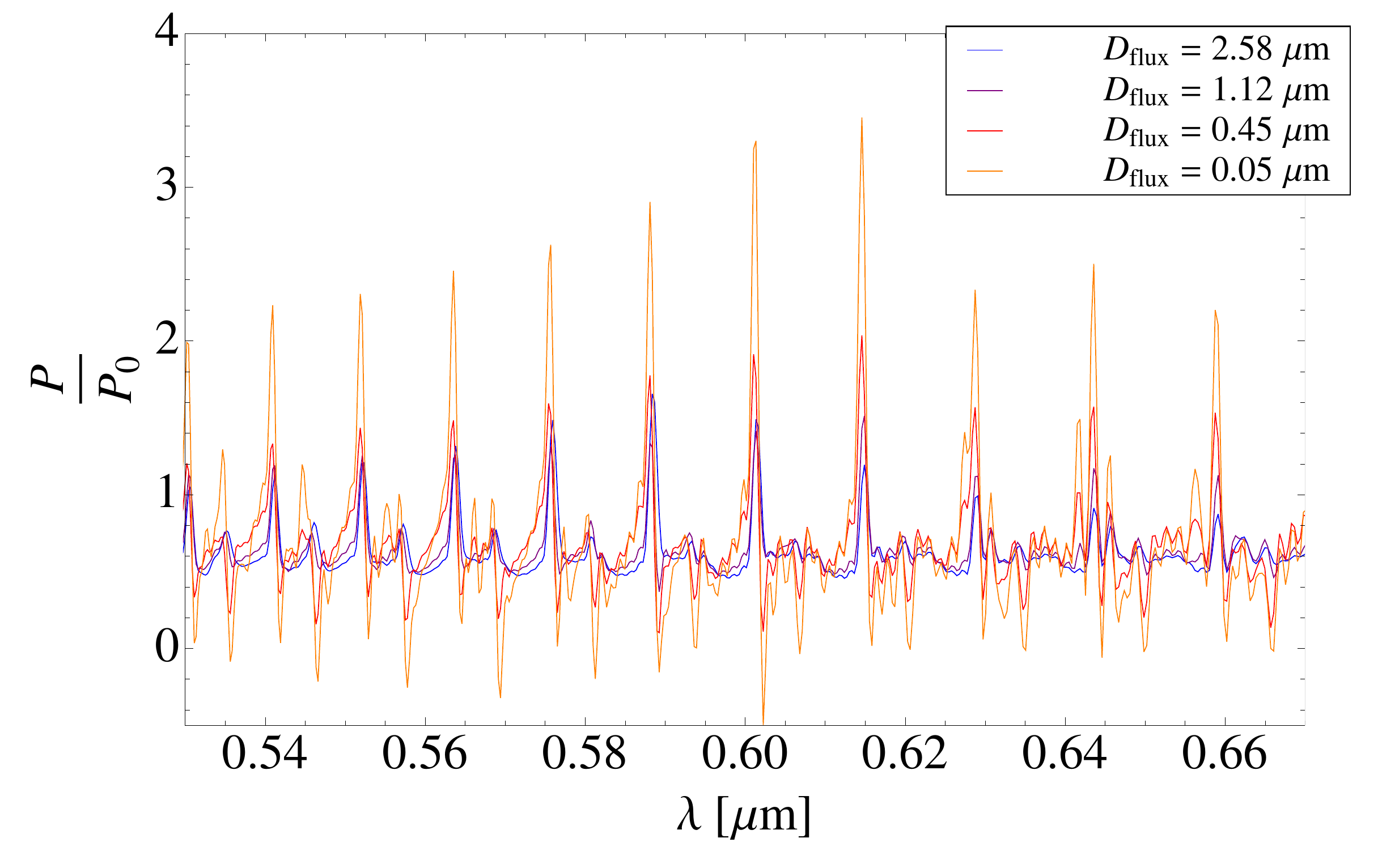}\\
\mbox{\hspace{3.3cm}(a)\hspace{6cm}(b)}
\vspace{-1mm}
\caption{\footnotesize{ A comparison of the power spectra 
of $6$ $\mu$m diameter microspheres with a tangential source ($\lambda = 0.6$ $\mu$m), (a) for 
flux collection regions at different distances $L_{\mathrm{flux}}$ 
from the centre of the sphere, and 
(b) for different flux region diameters $D_{\mathrm{flux}}$. A resolution of $\Delta x=22$ nm is used.}}
\label{fig:WGMang}
\end{figure}
\end{widetext}

The simulation can also provide information about the 
angular distribution of the flux, both directly and indirectly.
Since the FDTD simulation records the electromagnetic field values, the 
flux density, $\mathbf{S}(\mathbf{r})$, may be projected onto the circular 
region of the 
$z$-$y$ plane (with normal vector $\hat{\mathbf{n}}$) for any wavelength value. 
An analysis of this type 
is helpful for 
visualizing the distribution, distinguishing the azimuthal and radial 
modes at different time slices, and  identifying transient resonant 
features.
By integrating the flux over a collection time of $1.0$ ps, 
the distribution over the collection plane indicates the angular dependence of the mode 
at a particular wavelength. 
As an example, Fig.~\ref{fig:WGMrad} shows the flux distribution over the collection region, 
for a tangential dipole source, and a surrounding medium of air. The four panels 
display the four most prominent WGM peaks occurring in that wavelength region, corresponding to 
values of $0.576$, $0.588$, $0.601$ and $0.615$ $\mu$m, respectively. 
It is apparent that the modes at wavelengths of $0.601$ $\mu$m (TE$_{38,1}$) and $0.615$ $\mu$m (TE$_{37,1}$) 
exhibit a flux distribution with peaks spread out in the flux measuring region 
over a comparatively wide angle. 
The total power output for these modes is likely 
to vary significantly if the diameter of the region is reduced. 
In contrast, the mode occurring at $0.588$ $\mu$m (TE$_{39,1}$) has a more focused 
concentration of flux in the centre of the measuring region, corresponding to a narrower 
angular distribution of flux. The power output for this mode will be  
 consistent over a wider range of flux region diameters. 

As a consequence of the nontrivial angular distributions of the modes, 
changes in the power spectrum are also detected.  
Figure~\ref{fig:WGMang} shows the spectrum for the $6$ $\mu$m sphere 
for flux planes at different distances from the sphere (Fig.~\ref{fig:WGMang}(a)), 
and different 
 diameters (Fig.~\ref{fig:WGMang}(b)). In each case, a tangential electric dipole is used, 
and the collection time is held fixed at $1.0$ ps. 
By comparing the 
spectra from several flux collection plane sizes and positions, 
changes in the 
peak heights are observed, indicating the different angular behavior 
of the modes. Changes in the distance of the flux region, 
$L_{\mathrm{flux}}$, result in a similar mode structure, but the overall 
power output, especially for the dominant WGM peaks, 
changes as the flux is sampled differently 
in each region. Alterations in the diameter of the flux region 
results in more noticeable 
changes to the mode heights, demonstrating the highly angular-dependent nature 
of the relative contribution of each mode to the total flux.

These scenarios indicate that the 
FDTD simulation package represents a promising development toward a 
robust, customizable, predictive 
toolkit for simulating the whispering gallery modes of microspherical 
resonators. 
In this investigation, the FDTD tool has already provided guidance 
for future designs of resonators, which can be adapted for biosensing
 applications. 
For example, in Fig.~\ref{fig:WGM}, dominance of TE or TM modes 
is dependent on the alignment of the dipole source on the surface 
of the sphere. Microsphere sensors may be able to detect the 
orientation of an external biomolecule by recording the 
relative coupling strengths of TE to TM modes.

\section{Conclusion}
\vspace{-1.66mm}
This work presents the first easily customizable approach for simulating 
whispering gallery modes of microspheres, which is based on the FDTD 
method. The tool is easily 
customizable, and especially suited to investigating 
a wide variety of resonator scenarios. 
The FDTD approach is able to simulate changes to the power spectrum 
by measuring flux in the near-field region for any length of time, 
using a circular flux collection plane.
The comprehensive inclusion of all modes  
allows for more realistic simulations than is feasible  
for typical analytic models. 

In addition, different methods of mode excitation 
of  a  microsphere can be investigated, 
such as dipole source locations and alignments. 
Changes in the peak positions of the modes due to discretization effects 
can be quantified, and absorbed into the systematic uncertainty.

In practice, one can scan over a variety of parameters to gauge the 
effect on the profile of the collected power spectrum. 
Consequently, a wavelength and resonator combination can be sought 
systematically, and optical attributes of particular interest may be
 emphasized or 
suppressed. The optimization of the coupling efficiency to certain  
mode channels allows one to preselect desired resonator properties 
in a cost-effective way, 
leading to new design solutions that might not otherwise have been found. 

As a first example, the angular behavior of the energy flux 
was investigated for several sizes and distances of the collection region, 
leading to alterations in the coupling efficiencies to different modes. This 
is  
observed in the profile of the power spectrum. 
The magnitude of the flux density projected onto
 the collection disk was also mapped out 
for a range of collection times, allowing one to distinguish the 
distribution of the modes at different collection times.

The new simulation results were compared and contrasted to 
two different analytic models. 
The analytic models rely on assumptions of a perfectly smooth dielectric, 
with the total radiation output measured in the far-field region. 
It was found that the peak positions 
from the models matched the FDTD simulations well.

The computational toolkit presented in this paper 
represents the first step in establishing a realistic and easy-to-use 
simulator, 
which is important for facilitating a 
cost-effective approach to designing tailored optical resonators 
for biosensing.

\section*{Acknowledgments}
\vspace{-1.66mm}
The authors acknowledge the support of the ARC Georgina Sweet  
Laureate Fellowship (T.~M.~M.), and the ARC Centre for Nanoscale BioPhotonics. 
The authors also 
acknowledge eResearchSA for the use of supercomputing resources.

\end{document}